\documentclass[11pt]{article}
\usepackage[usenames]{color} %used for font color
\usepackage[cmex10]{amsmath}
\usepackage{amssymb}
\usepackage{amsthm}
\usepackage{txfonts}

\newtheorem{cor}{Corollary}
\newtheorem{defn}{Definition}

\newtheorem{thm}{Theorem}
\newtheorem{eg}{Example}
\newtheorem{prop}{Proposition}

\def\R{\mathbb{R}}

\def\Zeal{\mathop{\hbox{\mit Z\kern-.29em Z}}\nolimits}
\def\be{\begin{equation}}
\def\ee{\end{equation}}
\def\pf{\noindent {\bf Proof} \ }
\def\endpf{\hfill $\square$}

\def\l{\label}
\def\r{\ref}
\def\bs{\backslash}
\def\ra{\rightarrow}
\def\ba{\begin{eqnarray*}}
\def\ea{\end{eqnarray*}}

\def\sr{\stackrel}

\def\E{{\rm E}}
\def\P{{\rm Pr}}

\def\bfb{{\bf b}}

\def\bfh{{\bf h}}

\def\bft{{\bf t}}
\def\bfu{{\bf u}}
\def\bfv{{\bf v}}
\def\bfw{{\bf w}}
\def\bfx{{\bf x}}

\def\bfX{{\bf X}}

\def\calE{{\cal E}}

\def\calH{{\cal H}}

\def\calS{{\cal S}}

\def\"{{\prime\prime}}
\def\sfN{{\sf N}}
\def\Nbar{\overline{\sf N}}

\def\dim{{\rm dim}}

\def\bqy{\begin{eqnarray}}
\def\eqy{\end{eqnarray}}
\def\ol{\overline}

\def\GV'{{G^*(V')}}
\def\<{\langle}
\def\>{\rangle}
\def\rv{random variable}
\def\half{{1 \over 2}}

\def\aa{$\ddot{\rm a}$}
\def\uui{$\ddot{\it u}$}
\def\AA{$\ddot{\rm A}$}

\def\bit{\bibitem}

\def\CS{Cauchy-Schwarz inequality}

\usepackage{graphicx}

\begin{document}

\newcommand{\dqdef}{\mbox {$  \ \stackrel{def}{=}\    $}  }

\begin{titlepage}
\title{Inequalities Revisited}
\author{Raymond W. Yeung\thanks{Raymond Yeung is with 
Institute of Network Coding and Department of Information
Engineering,
The Chinese University of Hong Kong, N.T., Hong Kong.
Email: whyeung@ie.cuhk.edu.hk} 
}

\maketitle

\begin{center}
{\bf Abstract} 
\end{center}

\noindent
In the past over two decades, very fruitful results have been obtained in information theory in the study of the Shannon entropy. This study has led to the discovery of a new class of constraints on the Shannon entropy called non-Shannon-type inequalities.
Intimate connections between the Shannon entropy and different branches of mathematics including group theory, combinatorics, Kolmogorov complexity, probability, matrix theory, etc, have been established. All these discoveries were based on a formality introduced for constraints on the Shannon entropy, which suggested the possible existence of constraints that were not previously known. We assert that the same formality can be applied to inequalities beyond information theory. To illustrate the ideas, we revisit through the lens of this formality three fundamental inequalities in mathematics: the AM-GM inequality in algebra, Markov's inequality in probability theory, and the Cauchy-Scharwz inequality for inner product spaces. Applications of this formality have the potential of leading to the discovery of new inequalities and constraints in different branches of mathematics.

%In the literature of mathematics, there are three common usages of the term “inequality”. The first usage carries the original meaning of an inequality. The second usage carries a derived meaning of an inequality. The third usage specifies a constraint that is universally satisfied within a particular domain. Well-known examples of this usage include the Cauchy-Schwarz inequality for inner product spaces, the Hadarmad inequality for positive definite matrices, the Ingleton inequality for representable matroids, etc. Inspired by a geometrical framework for entropy inequalities first introduced in information theory, which has led to very fruitful results well beyond information theory (e.g., finite group theory, combinatorics, Kolmogorov complexity, probability, matrix theory, quantum mechanics), we extend this idea to other domains. As an illustration, we revisit the Cauchy-Scharwz inequality and discuss its tightness under this light. Applications of this idea has the potential of obtaining many fundamental results related to inequalities in different domains.%

\end{titlepage}

%\pagenumbering{arabic}

\section{Introduction}
\l{sec:intro}
Inequality defined on the ordered field of real numbers is one of the most fundamental concepts in mathematics. 
In particular, a {\em universally quantified} inequality is one that holds for all
members in the domain of discourse satisfying certain conditions. Examples are
\begin{enumerate}
\item
For all $x \in \mathbb{R}$, $e^x \ge 1+x$.
\item
For all $x \in \{ y \in \mathbb{R} : y \ge 5 \}$, $x \ge 4$.
\end{enumerate}
The latter may simply be written as ``If $x \ge 5$, then $x \ge 4$".
Essentially, any inequality in mathematics that bears a name is a universally quantified inequality, for example, the AM-GM inequality,
the Cauchy-Schwarz inequality, Jensen's inequality, and Minkowski's inequality. The list goes on and on.

The main contribution of this paper is to introduce a framework that gives a geometrical interpretation of such inequalities. This framework has its origin in information theory \cite{Shannon48}, specifically in the study of inequalities on the Shannon entropy (or simply entropy when there is no ambiguity) in the late 1990s \cite{Yeung97}. This study has led to the discovery of so-called {\em non-Shannon-type} entropy inequalities, namely inequalities on the Shannon entropy beyond what were known before then (collectively called {\em Shannon-type} entropy inequalities). Subsequently, intimate relations between the Shannon entropy and different branches of mathematics including group theory, combinatorics, Kolmogorov complexity, probability, matrix theory, etc, have been established. Inspired by this development, there has also been a wave of pursuit of new inequalities on the von Neumann entropy, a generalization of 
the Shannon entropy to the quantum case.

We assert that this geometrical framework which has led to very fruitful results in the study of entropy inequalities can also be applied to general universally quantified inequalities. In this paper, we will develop the concepts starting with a very simple example, and then apply the concepts to increasingly elaborate examples. The results are presented in a logical order instead of the chronological order that the results were obtained.

The rest of the paper is organized as follows. In Section~\r{sec:AMGM},
we discuss the well-known inequality of arithmetic and geometric means (AM-GM inequality) in algebra, and show that the AM-GM inequality completely characterizes the relation between the AM and GM of a finite collection of nonnegative numbers. In Section~\r{sec:Markov}, we discuss Markov's inequality in probability theory. We show that for a nonnegative random variable $T$ and a nonnegative value $c$, Markov's inequality essentially completely characterizes the relation between the two quantities $E[T]$ 
(expectation of $T$) and $\P \{ T \ge c \}$.
In Section~\r{sec:CS}, we discuss the \CS \ for real inner product spaces.
We show that for two vectors $\bfu, \bfv \in V$, where $V$ is a real inner
product space, the \CS \ completely characterizes the relation among the three
quantities $\langle \bfu, \bfu \rangle$, $\langle \bfv, \bfv \rangle$, and 
$\langle \bfu, \bfv \rangle$ if and only if $\dim(V)$ (the dimension of $V$) is at least 2.
Nevertheless, there exists no inequality on the quantities 
$\< \bfu, \bfu \>$, $\< \bfv, \bfv \>$, and $\< \bfu, \bfv \>$ that holds for all inner product space $V$ (regardless of the value of $\dim(V)$) which is not implied by the \CS. The results in Sections~\r{sec:AMGM} to \r{sec:CS}
are new to our knowledge. In Section~\r{sec:entropy}, we give an exposition
of the study on entropy inequalities in information theory since the late 1990s. We also briefly discuss the relations between the Shannon entropy and network coding,
conditional independence of random variables, finite groups, positive semi-definite matrices, Kolmogorov complexity, and quantum mechanics. The paper is concluded in Section~\r{sec:conclusion}.

\section{The AM-GM Inequality}
\l{sec:AMGM}
\def\AG{AM-GM inequality}
The inequality of arithmetic and geometric means, or the AM-GM inequality in brief, is elementary in algebra. 
%It states that for a finite list of nonnegative numbers, the arithmetic mean is at least equal to the geometric mean, and equality holds if and only if all the numbers in the set are the same.
In this section, we use this very simple inequality as an example to illustrate the concepts we will develop in this work. 

The arithmetic mean and geometric mean of a finite list of nonnegative numbers $x_1, x_2, \ldots, x_n$ are  
\[
\mbox{AM} = {1 \over n} \, (x_1 + \ldots + x_n)
\]
and 
\[
\mbox{GM} = \sqrt[n]{ \, x_1 \cdot \ldots \cdot x_n \, } ,
\]
respectively. 
The \AG \ says that
\be
\mbox{AM} \ge \mbox{GM}.
\l{AM-GM}
\ee
For a collection of
proofs of the \AG, we refer the reader to \cite{AMGM}. Throughout this 
section, we will use AM and GM to denote the arithmetic mean and the geometric
mean of some finite list of nonnegative numbers, respectively.

Traditionally, the AM-GM inequality is interpreted as either a lower bound on the AM or an upper bound on the GM. Here, we take a somewhere different view on the AM-GM
inequality that will be elaborated in the rest of the section.

For a finite list of nonnegative numbers, the AM and GM are quantities of interest, and 
we are interested in the relation between these two quantities. 
The \AG \ is a characterization of this relation.
Since $x_i \ge 0$ for all $i$, we immediately have 
${\rm AM} \ge 0$ and $\rm{GM} \ge 0$. These inequalities come directly from the 
setup of the problem.
In fact, from ${\rm GM} \ge 0$ and ${\rm AM} \ge {\rm GM}$, we can obtain
${\rm AM} \ge 0$. Therefore, ${\rm AM} \ge 0$ is redundant.

%Again let $a$ and $g$ be real numbers. If $a$ is equal to the AM of 
%some finite list of nonnegative numbers and $g$ is equal to the GM of 
%the same set of numbers, are there any constraints on $a$ and $g$ other than the 
%AM-GM inequality? We will refer to this as the {\em sufficiency} of the 
%AM-GM inequality. Specifically, if the answer to the above question is no, then
%we say that the AM-GM inequality is sufficient.

We now introduce a geometrical framework for understanding the relation between 
the quantities AM and GM. Let $a$ and $g$ be the coordinates of $\R^2$,
the 2-dimensional Euclidean space, where $a$ and $g$ correspond to AM and GM, 
respectively. 
Define the region
\[
\Upsilon = \left\{ \, (a,g) \in \R^2 : g \ge 0 \ \mbox{and} \ a \ge g \, \right\}.
\]
where in the above, $g \ge 0$ and $a \ge g$ correspond to ${\rm GM} \ge 0$
and ${\rm AM} \ge {\rm GM}$, respectively.
See Figure~\r{aiupnav} for an illustration of $\Upsilon$. 

\begin{figure}[t]
	\centering
	\includegraphics[width=3.5in]{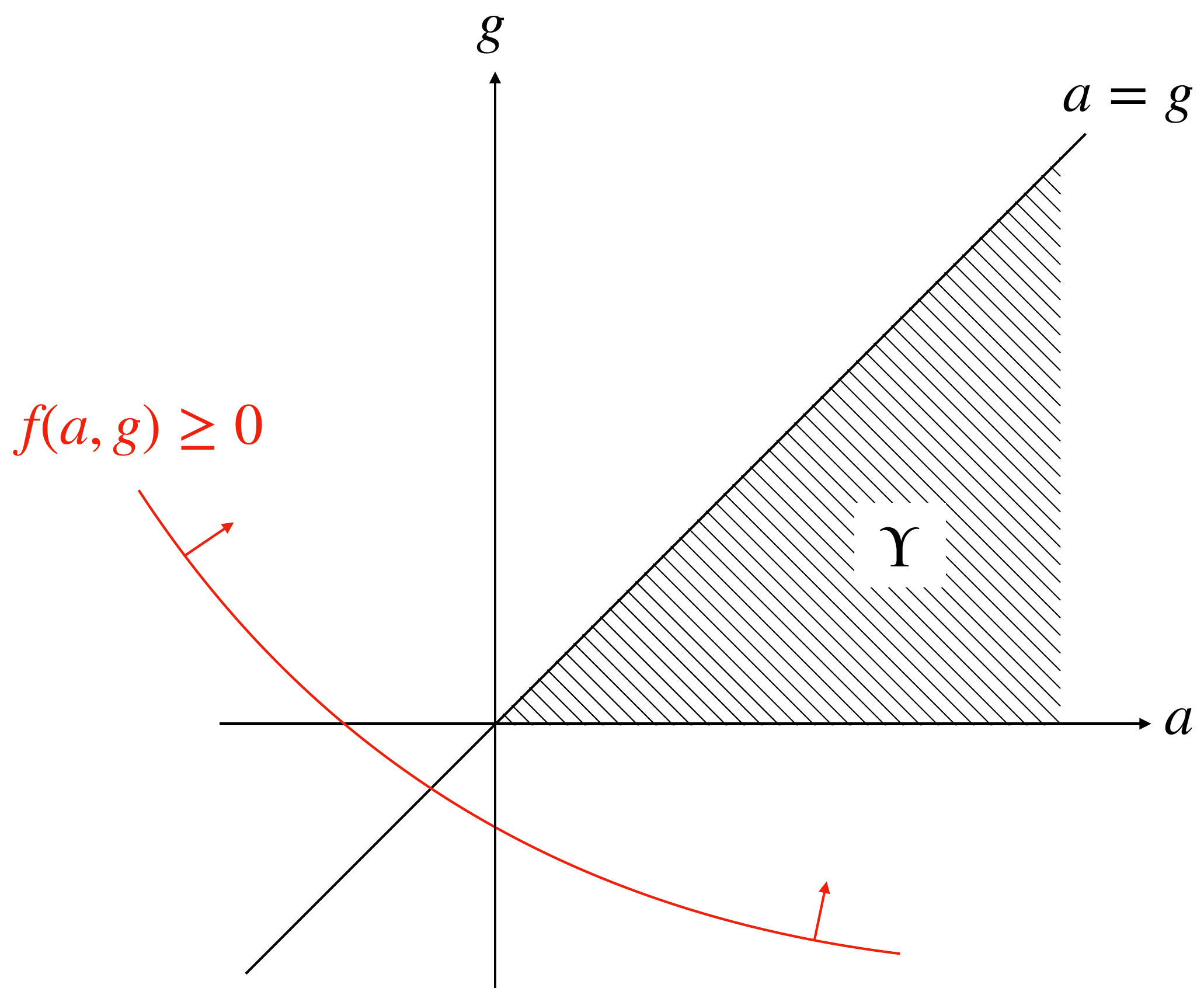}                                     
	\caption{The region $\Upsilon$ in $\mathbb{R}^2$ and an inequality $f(a,g) \ge 0$ implied by $g \ge 0$ and $a \ge g$.}
	\l{aiupnav}
\end{figure}

We now ask a very basic question: 
Are there constraints on AM and GM other than ${\rm GM} \ge 0$
and ${\rm AM} \ge {\rm GM}$?
As we will see, the answer to this question hinges on the next proposition.

\begin{prop}
For any $(a,g) \in \Upsilon$, there exist $x,y \ge 0$ such that 
\[
a = \frac{x+y}{2} \ \ \ \mbox{\rm and} \ \ \ g = \sqrt{xy} ,
\]
i.e., $a$ and $g$ are the AM and GM of the list of nonnegative numbers $x, y$, respectively.
\l{thm:xy}
\end{prop}

\pf \
Consider a fixed ordered pair $(a,g) \in \Upsilon$. Since $(a,g) \in \Upsilon$,
we have $g \ge 0$ and $a \ge g$.
Let
\be
a = \frac{x+y}{2} \ \ \ \mbox{\rm and} \ \ \ g = \sqrt{xy} ,
\l{qg9pha}
\ee
where $x, y \in \R$ are unknowns. Here, $x$ and $y$ can be obtained by solving the above simultaneous equations, which is elementary. Note that if $g < 0$, then
$g = \sqrt{xy}$ in the above cannot be satisfied, but this is not the case since $(a,g) \in \Upsilon$.
The solution to (\r{qg9pha}) is 
\[
x = a \pm \sqrt{a^2 - g^2} \ \ \ \mbox{and} \ \ \ y = a \mp \sqrt{a^2-g^2} ,
\]
where $a^2 - g^2 \ge 0$ since $a \ge g$. Therefore, $x$ and $y$ are always real.
Finally, it follows from $\sqrt{a^2 - g^2} \le a$ that $x$ and $y$ are always 
nonnegative. The proposition is proved.
\endpf

\bigskip
An ordered pair $(a,g)$ is an {\em achievable pair}, or simply achievable,\footnote{
The discussion in this section is built upon the concept of ``achievability", which will continue to play a central role in the rest of the paper.
We note that {\em achievability} is a fundamental concept  in information theory for formulating coding theorems. For example, C.~E.~Shannon's celebrated source coding theorem \cite{Shannon48} states that the coding rate for lossless data compression must be at least equal to the entropy rate of the information source. This means that any coding rate above the entropy rate of the information source is asymptotically achievable by some encoding-decoding schemes.
} if 
$a$ and $g$ are respectively the AM and GM of 
some finite list of nonnegative numbers.
The next theorem is a consequence of Proposition~\r{thm:xy}.

\begin{thm}
An ordered pair $(a,g) \in \R^2$ is achievable if and only if $(a,g) \in \Upsilon$.
\l{prop:achievable}
\end{thm}

\proof
Proposition~\r{thm:xy} implies that every ordered pair $(a,g) \in \Upsilon$ is 
achievable. On the other hand, if an ordered pair $(a,g)$ is achievable,
then $(a,g) = ({\rm AM}, {\rm GM})$ where AM and GM are respectively the arithmetic mean and
geometric mean of some finite list of nonnegative numbers. Therefore,
$g = {\rm GM} \ge 0$ and $a = {\rm AM} \ge {\rm GM} = g$, i.e., $g \ge 0$ and
$a \ge g$, implying that $(a,g) \in \Upsilon$. 
Hence, an ordered pair $(a,g)$ is achievable if and only if $(a,g) \in \Upsilon$. The theorem is proved.
\endpf

\bigskip
Theorem~\r{prop:achievable} says that $\Upsilon$ is precisely the set of 
all achievable pairs, thus completely characterizing the relation 
between the arithmetic and geometric means of a finite list of 
nonnegative numbers.
Since $\Upsilon$ is defined by $g \ge 0$ (corresponding to ${\rm GM} \ge 0$) and $a \ge g$ (corresponding to the AM-GM inequality), where the former comes directly from the setup of
the problem and can be regarded as given, we say that 
the AM-GM inequality completely characterizes the relation 
between the AM and GM of a finite list of nonnegative numbers.

An inequality in the quantities AM and GM has the general form
\be
f({\rm AM}, {\rm GM}) \ge 0 ,
\l{qgquph}
\ee
where $f : \R^2 \ra \R$.\footnote{
For any $A \subset \R^2$, if we let
\[
f(a,g) = \left\{ \begin{array}{ll}
1 & \mbox{if $(a.g) \in A$} \\
-1 & \mbox{if $(a.g) \not\in A$},
\end{array} \right.
\]
then $(a,g) \in A$ if and only if $f(a,g) \ge 0$.
Thus an inequality of the form (\r{qgquph}) can constrain
$({\rm AM}, {\rm GM})$ to any subset of $\R^2$.
} 
For example, if $f(a,g) = a-g$, then 
(\r{qgquph}) becomes the AM-GM inequality. Let
\[
R_f = \left\{ \, (a,g) \in \R^2 : f(a,g) \ge 0 \, \right\}
\]
be the region in $\R^2$ induced by $f \ge 0$.
If (\r{qgquph}) is satisfied for all finite lists of nonegative numbers, 
we say that the inequality is {\em valid}.

\begin{thm}
The inequality (\r{qgquph}) is valid if and only if 
\be
\Upsilon \subset R_f.
\l{qtrehq2}
\ee 
\l{prop:inclusion}
\end{thm}

\proof
We first prove the ``if" part. Assume that (\r{qtrehq2}) holds.
Consider any finite list of nonnegative numbers and let AM and GM be the
arithmetic mean and geometric mean, respectively. Then $({\rm AM}, {\rm GM})$
is achievable, and so by Theorem~\r{prop:achievable}, $({\rm AM}, {\rm GM}) \in \Upsilon$. Then by (\r{qtrehq2}),
$({\rm AM}, {\rm GM}) \in R_f$, implying that $f({\rm AM}, {\rm GM}) \ge 0$.
This shows that the inequality (\r{qgquph}) is valid.

Next, we prove the ``only if" part by contradiction. Assume that the inequality
(\r{qgquph}) is valid, i.e., 
$f({\rm AM}, {\rm GM}) \ge 0$ is satisfied by all finite lists of nonnegative numbers, but
\[
\Upsilon \not\subset \left\{ \, (a,g) \in \R^2 : f(a,g) \ge 0 \, \right\}.
\]
Then there exists an ordered pair $(a_0, g_0) \in \Upsilon$ such that $f(a_0,g_0) < 0$.
Since $(a_0, g_0) \in \Upsilon$, by Theorem~\r{prop:achievable}, $(a_0,g_0)$ is achievable, which means that $(a_0, g_0) = ({\rm AM}^*, {\rm GM}^*)$, where AM$^*$ and GM$^*$ are respectively the arithmetic mean and geometric mean of some finite list of nonnegative numbers. In other words, we have 
$f({\rm AM}^*, {\rm GM}^*) = f(a_0, g_0) < 0$, which is a contradiction to the assumption that 
$f({\rm AM}, {\rm GM}) \ge 0$ is satisfied by all finite lists of nonnegative numbers. The theorem is proved.
\endpf

\bigskip
Theorem~\r{prop:inclusion} gives a complete characterization of all valid 
inequalities in AM and GM. Specifically, 
$f({\rm AM}, {\rm GM})) \ge 0$ is a valid inequality in AM and GM if and only if 
$R_f$,
the region induced by $f \ge 0$, is an outer bound on $\Upsilon$.
This is illustrated in Figure~\r{aiupnav}.

In Theorem~\r{prop:inclusion}, the set inclusion in (\r{qtrehq2}) is equivalent to
\[
\left. \begin{array}{r}
g \ge 0 \\
a \ge g
\end{array} \right\}
\Rightarrow f(a,g) \ge 0 .
\]
Upon replacing the dummy variables $a$ and $g$ by AM and GM, respectively, the above becomes 
\be
\left. \begin{array}{r}
{\rm GM} \ge 0 \\
{\rm AM} \ge {\rm GM}
\end{array} \right\}
\Rightarrow f({\rm AM},{\rm GM}) \ge 0 ,
\l{qt8hfaf}
\ee
meaning that any valid inequality in AM and GM is implied by the inequalities $\mbox{GM} \ge 0$ and $\mbox{AM} \ge \mbox{GM}$.
Hence, we conclude that there exists no inequality in AM and GM other than these two inequalities.
As discussed, since the inequality ${\rm GM} \ge 0$ comes directly from the setup of the problem and can be regarded as given, in view of (\r{qt8hfaf}), we say that the AM-GM inequality is {\em sharp}.

We end this section with a remark on the {\em tightness} of a valid inequality. 
Let $f({\rm AM}, {\rm GM}) \ge 0$ be a valid inequality.
By Theorem~\r{prop:inclusion}, the set inclusion in (\r{qtrehq2}) holds. 
If $f({\rm AM}, {\rm GM}) \ge 0$ is tight, then $f({\rm AM}^*, {\rm GM}^*)=0$ for some achievable pair $({\rm AM}^*, {\rm GM}^*)$, which by 
Theorem~\r{prop:achievable} is in $\Upsilon$. If the boundary of the region $R_f$ is equal to 
\[
\left\{ \, (a,g) \in \R^2 : f(a,g) =0 \, \right\},
\]
then $({\rm AM}^*, {\rm GM}^*)$ is in $\Upsilon$ as well as on the boundary of $R_f$. This implies that the boundary of $R_f$ touches the region $\Upsilon$
at the point where $f \ge 0$ is tight, providing a geometrical interpretation of the tightness of a valid inequality.

As an example,
the inequality $2 {\rm AM} \ge {\rm GM}$ is equivalent to $f({\rm AM}, {\rm GM}) \ge 0$ with $f(a,g) = 2a - g$. It is a valid inequality because $\Upsilon \subset R_f$. Moreover, $2 {\rm AM} \ge {\rm GM}$ is tight for the list of nonnegative number, 0, with $({\rm AM}^*, {\rm GM}^*) = (0,0)$.
Accordingly, the boundary of $R_f$, namely the set
\[
\left\{ \, (a,g) \in \R^2 : 2a = g \, \right\},
\]
touches the region $\Upsilon$ at the origin.
This is illustrated in Figure~\r{touch}. 
%Although the inequality 
%$2 {\rm AM} \ge {\rm GM}$ is tight, it is not sharp as to be discussed in the next remark.

%\item 
%Let $f_\Upsilon : \R^2 \ra \R$ be such that $(a,g) \in \Upsilon$ if and only if $f_\Upsilon(a,g) \ge 0$. For example, we can let
%\[
%f_\Upsilon(a,g) = \left\{ \begin{array}{ll}
%1 & \mbox{if $(a,g) \in \Upsilon$} \\
%-1 & \mbox{if $(a,g) \not\in \Upsilon$.}
%\end{array}
%\right.
%\]
%Evidently, there is more than one way to define $f_\Upsilon$ for this purpose.
%Theorem~\r{prop:inclusion} says that $f({\rm AM}, {\rm GM}) \ge 0$ is a valid inequality in AM and GM if and only if $\Upsilon \subset R_f$, or equivalently,
%\[
%f_\Upsilon({\rm AM}, {\rm GM}) \ge 0 \ \ \ \Rightarrow \ \ \ f({\rm AM}, {\rm GM}) \ge 0.
%\]
%Therefore, $f_\Upsilon({\rm AM}, {\rm GM}) \ge 0$ is the {\em sharpest possible} inequality in AM and GM. 

%Evidently, $2 {\rm AM} \ge {\rm GM}$ does not imply $f_\Upsilon({\rm AM}, {\rm GM}) \ge 0$. For example, 

\begin{figure}[t]
	\centering
	\includegraphics[width=3.5in]{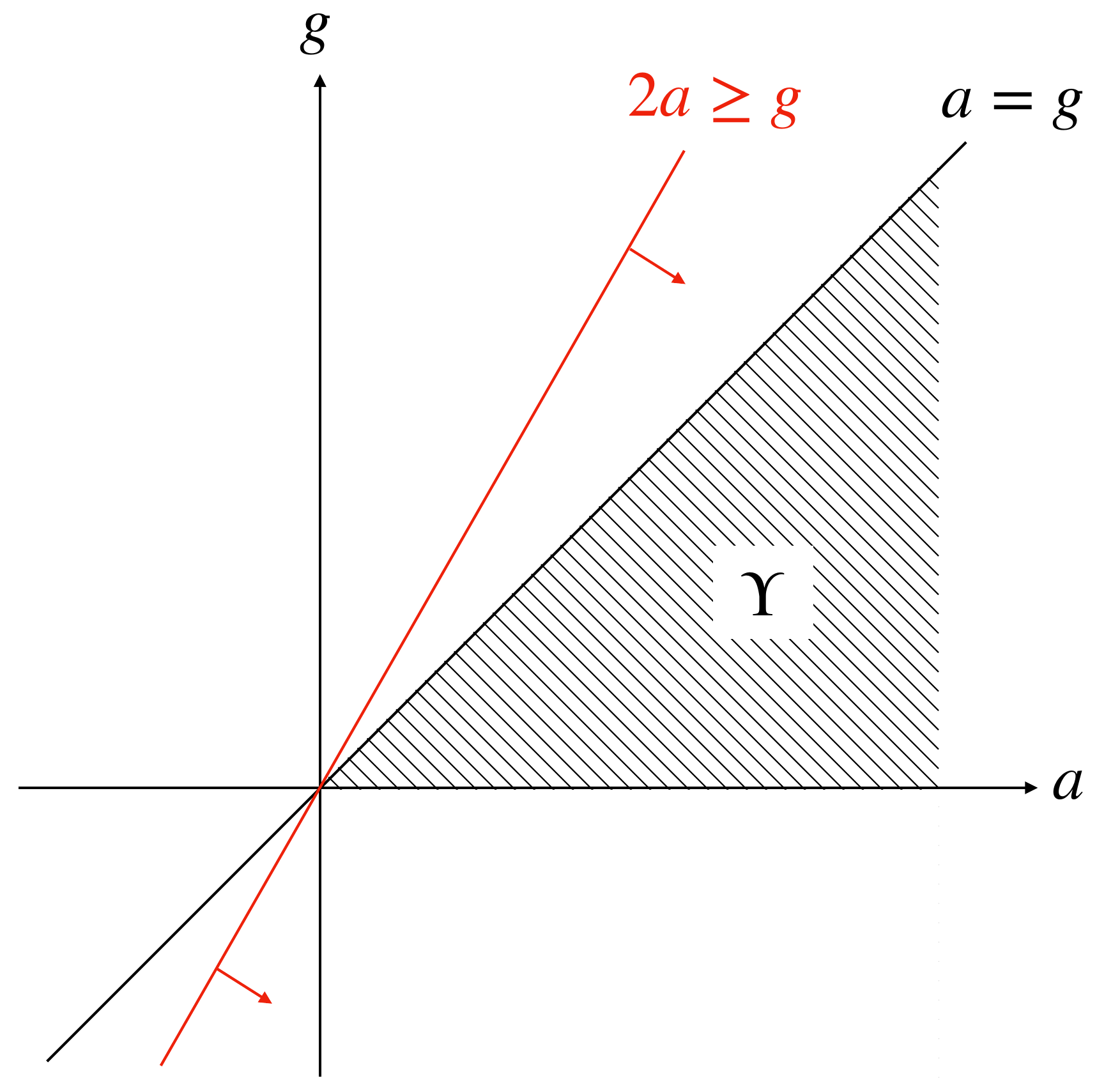}                                     
	\caption{An illustration of the region $R_f$ for $f = 2a - g$.}
	\l{touch}
\end{figure}

%\end{enumerate}

%Before ending this section, we remark that an inequality
%$f \ge 0$, where $f: \R^m \ra \R$, can represents the most general form of constraints in $\R^m$. Let $S$ be any subset of $\R^m$. By letting
%\[
%f(x) = \left\{ \begin{array}{ll}
%1 & \mbox{if $f(x) \in S$} \\
%-1 & \mbox{otherwise},
%\end{array}
%\right.
%\]
%then the set 
%\[
%\left\{ \, x \in \R^m : f(x) \ge 0 \, \right\}
%\]
%is exactly equal to $S$. Therefore, an inequality of the form $f \ge 0$ can 
%constrain $x$ to any subset of $\R^m$.

\section{Markov's Inequality}
\l{sec:Markov}
In probability theory, Markov's inequality asserts that for a nonnegative \rv \ $T$ and any fixed $c > 0$, 
\be
\P \{ T \ge c \} \le \frac{\E [T]}{c} ,
\l{Markov}
\ee
where $\E [T]$ denotes the expectation of $T$. Here,
$\P \{ T \ge c \}$ and $E[T]$ are two quantities of interest for any nonnegative
random variable $T$, and we are interested in the relation between them.
Markov's inequality gives a characterization of this relation.

Let $c$ in (\r{Markov}) be fixed, and let $F_T$ denote the probability distribution of $T$, i.e., $F_T(t) = \P \{ T \le t \}$. The following is a proof of (\r{Markov}):
\ba
\E [T]
& = & \int_{t \ge 0} t \, dF_T(t) \\
& = & \int_{0 \le t < c} t \, dF_T(t) + \int_{t \ge c} t \, dF_T(t) \\
& \sr{\rm i)}{\ge} & \int_{t \ge c} t \, dF_T(t) \\
& \sr{\rm ii)}{\ge} & c \int_{t \ge c} dF_T(t) \\
& = & c \, \P \{ T \ge c \} .
\ea
Then (\r{Markov}) is obtained by dividing both sides about by $c$.
In the above, the inequality i) is tight if and only if
\be
\int_{0 \le t < c} t \, dF_T(t) = 0 ,
\l{5geqrp9n}
\ee
or 
\[
\P \{ 0 < T < c \} = 0.
\]
Note that although $\P \{ T = 0 \}$ can be strictly positive, it would not 
in any case make any 
contribution to (\r{5geqrp9n}).
On the other hand, the inequality ii) is tight if and only if 
\[
\P \{ T > c \} = 0,
\]
or equivalently,
\be
\P \{ T \ge c \} = \P \{ T = c \}.
\l{oia4hoin}
\ee
If (\r{Markov}) is tight, i.e., i) and ii)
are tight simultaneously, then $F_T$ can only have two point masses, one at 0 and the other at $c$, with 
\[
\P \{ T = 0 \} + \P \{ T = c \} = 1.
\]
As a sanity check, for this distribution, from (\r{oia4hoin}), we have 
\[
E[T] = c \cdot \P \{ T = c \} = c \cdot \P \{ T \ge c \} ,
\]
or
\be
\P \{ T \ge c \} = \frac{E[T]}{c} .
\l{Markov=}
\ee
Thus (\r{Markov}) indeed holds with equality.

Note that the above discussion holds regardless of the value of $\P \{ T \ge c \}$
(which can be any number between 0 and 1). Thus Markov's inequality can hold with equality for 
all values of $\P \{ X \ge c \}$.

We can obtain further insight on
Markov's inequality by means of a geometrical framework similar to the one
discussed in Section~\r{sec:AMGM} for the AM-GM inequality. Continue to assume that $c > 0$ is fixed.
Let $p$ and $m$ be real numbers such that $p = \P \{ T \ge c \}$ and $m = E[T]$
for some nonnegative random variable $T$. Then from (\r{Markov}), we have
$m \ge c p$. Since $p$ is a probability, we also have
$0 \le p \le 1$. Now regard $p$ and $m$ as the two coordinates in $\R^2$, 
and define the region
\[
\Psi_c = \left\{ \, (p,m) \in \R^2 : 0 \le p \le 1 \ \mbox{and} \ m \ge c p \, \right\} .
\]
See Figure~\r{fig_pm}.

\begin{figure}[t]
\centering
\includegraphics[width=3.5in]{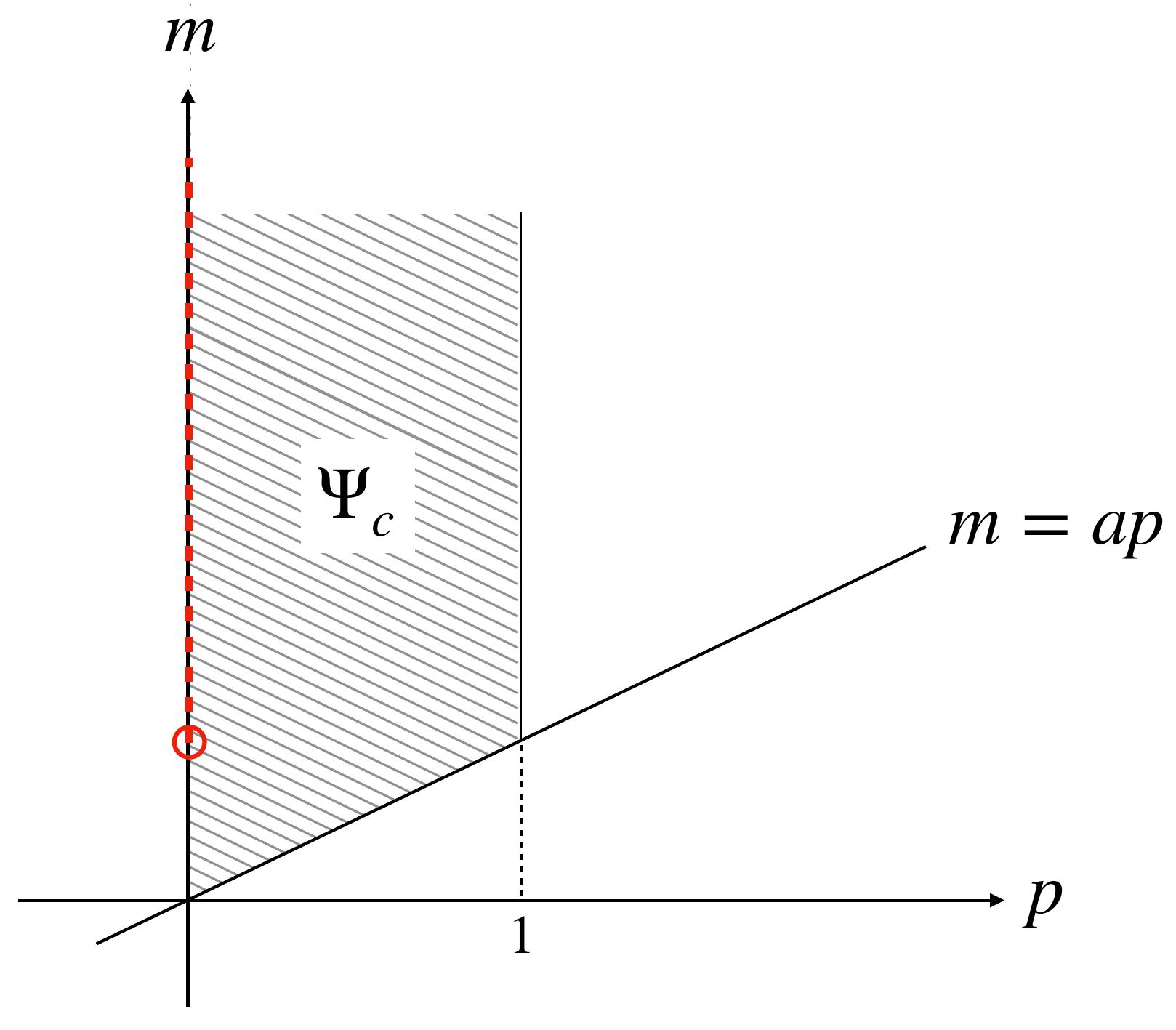}
\caption{The region $\Psi_c$ in $\mathbb{R}^2$.}
\l{fig_pm}
\end{figure}

For real numbers $p$ and $m$, if there exists
a nonnegative random variable $T$ such that $p = \P \{ T \ge c \}$ and
$m = E[T]$, we say that the ordered pair $(p,m) \in \R^2$ is {\em achievable} (by the random variable $T$). 
Note that an ordered pair 
in $\R^2$ may be achievable by more than
one nonnegative random variable, i.e., 
\[
(p, m) = (\P \{ T \ge a\}, E[T]) = (\P \{ T' \ge a\}, E[T'])
\]
where $T$ and $T'$ have different probability distributions. 

We will show that almost all ordered pairs in $\Psi_c$ are achievable.
We first define the {\em achievable region}
\[
\Psi_c^* = \left\{ \, (p,m) \in \R^2 : (p,m) \ \mbox{is achievable} \, \right\} .
\]
Since for any nonnegative random variable $T$, $0 \le \P \{ T \ge c \} \le 1$ and
Markov's inequality is satisfied, if $(p,m)$ is achievable, then $(p,m) \in \Psi_c$. Consequently, $\Psi_c^* \subset \Psi_c$.

\begin{thm}
$\Psi_c^* = \Psi_c \, \bs \, \{ (0,m) : m \ge c \}$.
\end{thm}

\proof
First, we prove that $\Psi_c^* \subset \Psi_c \, \bs \, \{ (0,m) : m \ge c \}$.
Since $\Psi_c^* \subset \Psi_c$, we only need to show that $(0,m)$ is not achievable for all $m \ge c$. When $p=0$, we have $\P \{ T \ge c \} = p = 0$, or 
$\P \{ 0 \le T < c \} = 1$. This implies that $m = E[T] < c$. In other words, every ordered pair 
$(0, m)$ where $m \ge c$ is not achievable.

Now, we prove that $\Psi_c \, \bs \, \{ (0,m) : m \ge c \} \subset 
\Psi_c^*$. 
For any $0 < p \le 1$, consider any $(p,m) \in \Psi_c$ and the following probability mass function for a \rv \ $T$:
\[
\P \{ T = 0 \} = 1-p \ \ \ \mbox{and} \ \ \ \P \{ T = m/p \} = p .
\]
Since $m \ge cp$, we have $m/p \ge c$, and so 
\[
\P \{ T \ge c \} = \P \{ T = m/p \} = p ,
\]
and $E[T] = p \, (m/p) = m$. Therefore, $(p,m)$ is achieved by the random variable 
$T$ as constructed. Note that unless $m = cp$, $(p,m)$ can always be achieved by more than one probability distribution.

It remains to prove that every ordered pair $(0,m)$ with $0 \le m < c$ is achievable. This can be done by noting that such an ordered pair can be achieved by any random variable $T$ with $\P \{ T = m \} = 1$. The theorem is proved.
\endpf

\bigskip
The region $\Psi_c$ is defined by Markov's inequality together with the constraint
$0 \le p \le 1$ which comes from the setup of the problem, and we have shown that
for any fixed $c > 0$, every ordered pair in $\Psi_c$ except for a region with Lebesgue measure 0 (namely the region $\{ (0,m) : m \ge c \}$) is achievable by some random variable $T$. Specifically:
\begin{itemize}
\item 
When $\P \{ T \ge c \} > 0$, Markov's inequality, namely
\be
E[T] \ge c \cdot \P \{ T \ge c \} ,
\l{t9phear}
\ee
which gives a lower bound on $E[T]$, is the only
constraint on $E[T]$ in terms of $\P \{ T \ge c \}$.
\item 
When $\P \{ T \ge c \} = 0$, Markov's inequality as in (\r{t9phear}), which becomes 
$E[T] \ge 0$, continues to be valid. However, we also have
\be
E[T] < c.
\l{wrw9io}
\ee
Combining (\r{t9phear}) and (\r{wrw9io}), we have 
\[
0 \le E[T] < c .
\]
\end{itemize}
See Figure~\r{fig_pm} for an illustration.
Since every ordered pair in $\Psi_c$ except for a region with Lebesgue measure 0 is achievable, or equivalently, $\ol{\Psi^*_c} = \Psi_c$, we say that Markov's inequality almost completely characterizes
the relation between $\P \{ T \ge c \}$ and $E[T]$.

Now consider an inequality in the quantities $\P \{ T \ge c \}$ and $E[T]$:
\be
f(\P \{ T \ge c \}, E[T]) \ge 0 ,
\l{q9g8hq}
\ee
where $f: \R^2 \ra \R$. For example, if $f(m,p) = m-cp$, then the inequality in (\r{q9g8hq}) becomes (\r{Markov}), namely Markov's inequality.
Let
\[
R_f = \left\{ \, (p,m) \in \R^2 : f(p,m) \ge 0 \, \right\}
\]
be the region in $\R^2$ induced by $f \ge 0$. 

%We note that $f(p,m) \ge 0$ is the most general form of
%a constraint on $(p,m) \in \R^2$. To see that, 
%let $A$ be an arbitrary subset of $\R^2$ (an arbitrary region in $\R^2$). By letting 
%\[
%f_A(p,m) = \left\{ \begin{array}{ll}
%1 & \mbox{if $(p,m) \in A$} \\
%-1 & \mbox{if $(p,m) \not\in A$},
%\end{array} \right.
%\]
%we have
%\ba
%R_{f_A} & = & \left\{ \, (p,m) \in \R^2 : f_A(p,m) \ge 0 \, \right\} \\
%& = & \left\{ \, (p,m) \in \R^2 : f_A(p,m) =1 \, \right\} \\
%& = & \left\{ \, (p,m) \in \R^2 : (p,m) \in A \, \right\} \\
%& = & A.
%\ea
%Therefore, an inequality $f(p,m) \ge 0$ can constrain $(p.m)$ to 
%any subset of $\R^2$.

If (\r{q9g8hq}) holds for all nonnegative random variable $T$, we say that the inequality is {\em valid}. 
Markov's inequality is such an example.
The fundamental importance of the achievable region $\Psi_c^*$ is explained in the next theorem, 
which asserts that an inequality $f(\P \{ T \ge c \}, E[T]) \ge 0$ is valid if and only if $R_f$ is an outer bound on $\Psi_c^*$. This implies that the region $\Psi_c^*$ completely characterizes all valid inequalities of the form (\r{q9g8hq}).

\begin{thm}
	The inequality (\r{q9g8hq}) is valid if and only if \ $\Psi_c^* \subset R_f$.
	\l{prop:subset}
\end{thm}

\proof
We first show that if $\Psi_c^* \subset R_f$, then (\r{q9g8hq}) holds for all nonnegative random variable $T$. Assume that $\Psi_c^* \subset R_f$. Consider any nonnegative random variable $T$ and the ordered pair $(\P \{ T \ge c\}, E[T])$. By the definition of $\Psi_c^*$, we have $(\P \{ T \ge c\}, E[T]) \in \Psi_c^* \subset R_f$, so that $(\P \{ T \ge c\}, E[T]) \in  R_f$.
It then follows from the definition of $R_f$ that $f(\P \{ T \ge c\}, E[T]) \ge 0$.

Next, we show that if (\r{q9g8hq}) holds for all nonnegative random variable $T$, then $\Psi_c^* \subset R_F.$
Consider any $(p,m) \in \Psi_c^*$. Then $(p,m) = (\P \{ T \ge c \}, E[T])$ for some
nonnegative random variable $T$. Since by our assumption (\r{q9g8hq}) holds for all 
nonnegative random variable $T$, we see that $(p,m) 
\in R_f$, and hence $\Psi_c^* \subset R_f.$

%Now we show that if $\Psi^* \not\subset R_f$, then there exists a nonnegative random variable $T$ such that (\r{q9g8hq}) does not hold.
%If $\Psi^* \not\subset R_f$, then there exists $(p,m)$ in $\Psi^*$ but not in $R_f$. Since $(p,m) \in \Psi^*$, $(p,m) = (\P \{ T \ge a\}, E[T])$ for some nonnegative random variable $T$.
%Since $(p,m) \not\in R_f$, the inequality (\r{q9g8hq}) does not hold for $T$.

Combining the above, we conclude that $\Psi_c^* \subset R_f$ is a necessary and sufficient condition for the inequality set (\r{q9g8hq}) to be valid. The theorem is proved.
\endpf

\bigskip
We now consider a finite set of inequalities on $\P \{ T \ge c\}$ and $E[T]$, 
\be
F = \left\{ \, f_i(\P \{ T \ge c\}, E[T]) \ge 0, 1 \le i \le k \, \right\} ,
\l{atqpvioa}
\ee
where 
$f_i: \R^2 \ra \R$. We say that $F$ is valid if $f_i(\P \{ T \ge c\}, E[T]) \ge 0$ is valid for all $i$.
Let
\[
R_F = \left\{ \, (p,m) \in \R^2 : f_i(p,m) \ge 0, 1 \le i \le m \,
\right\}
\]
be the region in $\R^2$ induced by $F$. The following is a corollary of
Theorem~\r{prop:subset}.

\begin{cor}
The set of inequalities (\r{atqpvioa}) is valid if and only if $\Psi_c^* \subset R_F$.
\l{cor:subset}
\end{cor}

\proof
Let 
\[
\tilde{f}(p,m) = \left\{ \begin{array}{ll}
	1 & \mbox{if $f_i(p,m) \ge 0$ for all $i$} \\
	-1 & \mbox{otherwise.}
\end{array} \right.
\]
Then $\tilde{f}(p,m) \ge 0$ if and only if $f_i(p,m) \ge 0$ for all $i$. In other words,
the set of inequalities (\r{atqpvioa}) is valid if and only if $\tilde{f}(p,m) \ge 0$ is valid, and hence $R_F = R_{\tilde{f}}$. Then the corollary is proved by applying Thoerem~\r{prop:subset}.
\endpf

\bigskip
We end this section with an example for Corollary~\r{cor:subset}.

\begin{eg}
Let 
\[
f_1 = p, \ \ f_2 = 1-p, \ \  f_3 = m-cp ,
\]
and $F = \{ \, f_i(\P \{ T \ge c\}, E[T]) \ge 0, i = 1, 2, 3 \, \}$. Then $R_F$ becomes $\Psi_c$. Since 
$f_i(\P \{ T \ge c\}, E[T]) \ge 0$ is valid for $i=1,2,3$, so is the inequality set $F$. Then by Corollary~\r{cor:subset}, $\Psi_c^* \subset \Psi_F$, which is indeed the case.
\end{eg}

\section{Cauchy-Schwarz Inequality}
\l{sec:CS}
The Cauchy-Schwarz inequality, which applies to a general inner product space, is among the 
most important inequalities in mathematics. For the purpose of this work, 
it suffices to confine our discussion to {\em real} inner product spaces.

\begin{defn}
A real inner product space is a vector space $V$ over $\mathbb{R}$ together with an inner product $\langle \cdot, \cdot \rangle : V \times V \ra \mathbb{R}$ that satisfies
the following for any vectors $\bfu, \bfv, \bfw \in V$ and any scalars $a, b \in \R$:
\begin{enumerate}
\item 
(Symmetry) $\langle {\bf u}, {\bf v} \rangle = \langle {\bf v}, {\bf u} \rangle$
\item 
(Linearity) $\langle a {\bf u} + b{\bf v}, {\bf w} \rangle = a \langle {\bf u}, {\bf w} \rangle + b \langle {\bf v}, {\bf w} \rangle$
\item 
(Positive-definiteness) $\langle {\bf u}, {\bf u} \rangle > 0$ if ${\bf u} \ne 0$.
\end{enumerate}
\end{defn}
The \CS \ asserts that for any $\bfu, \bfv \in V$, 
\be
\langle {\bf u}, {\bf v} \rangle^2 \le
\langle {\bf u}, {\bf u} \rangle \, \langle {\bf v}, {\bf v} \rangle ,
\l{Cauchy}
\ee
with equality if and only if $\bfu$ and $\bfv$ are linearly dependent. Here, 
for any pair of vectors $\bfu$ and $\bfv$, the three quantities of interest
are $\langle \bfu, \bfu \rangle$, $\langle \bfv, \bfv \rangle$, and 
$\langle \bfu, \bfv \rangle$, and we are interested in the relation among them.
The \CS \ gives a characterization of this relation.
In this section, we discuss this inequality in the spirit of the discussion in the previous sections.

In view of the three quantities involved in the \CS,
namely $\< \bfu, \bfu \>$, $\< \bfv, \bfv \>$, and $\< \bfu, \bfv \>$, we are motivated to consider the ordered triple $(\< \bfu, \bfu \>, \< \bfv, \bfv \>, \< \bfu, \bfv \> )$ for any $\bfu, \bfv \in V$.
For $(x,y,z) \in \R^3$, if $(x,y,z) = (\< \bfu, \bfu \>, \< \bfv, \bfv \>, \< \bfu, \bfv \> )$ for some
$\bfu, \bfv \in V$, then we say that $(x,y,z)$ is {\em achievable}.
We then define the {\em achievable region}
\[
\Phi^* = \left\{ \ (x, y, z) \in \R^3 : (x, y, z) \ \mbox{is achievable} \ \right\}.
\]
Also note that an ordered triple 
in $\R^3$ may be achievable by more than
one pair of vectors, i.e., 
\[
(x, y, z) = (\< \bfu, \bfu \> , \< \bfv, \bfv \>, \< \bfu, \bfv \> ) = (\< \bfu', \bfu' \> , \< \bfv', \bfv' \>, \< \bfu', \bfv' \>)
\]
where $(\bfu, \bfv) \ne (\bfu', \bfv')$. 
%As we will see, the achievable region $\Phi^*$ is of core interest in discussing the tightness of the \CS.

Let $V$ be a fixed inner product space. Consider a finite set of inequalities on the quantities $\< \bfu, \bfu \>$, $\< \bfv, \bfv \>$, and $\< \bfu, \bfv \>$:
\be
F = \{ \, f_i(\< \bfu, \bfu \>, \< \bfv, \bfv \>, \< \bfu, \bfv \>) \ge 0: 1 \le i \le m \, \}
\l{q9g8hq1}
\ee
where $f_i: \R^3 \ra \R$. For example, if $f_i \, (x,y,z) = xy - z^2$, then the $i$th inequality in (\r{q9g8hq1}) becomes (\r{Cauchy}), the \CS.
Let
\[
R_F = \left\{ \, (x,y,z) \in \R^3 : f_i \, (x,y,z) \ge 0, 1 \le i \le m \, \right\}
\]
be the region in $\R^3$ induced by $F$. 

If an inequality in (\r{q9g8hq1}) holds for all $\bfu, \bfv \in V$, we say that the inequality is {\em valid}. 
The \CS \ is such an example.
If the inequality in (\r{q9g8hq1}) is valid for all $i$, we say that the inequality 
set $F$ is valid.
The next theorem on the fundamental importance of the achievable region $\Phi^*$ follows directly from the discussion in Section~\r{sec:Markov}, so the proof is
omitted here.

\begin{thm}
For any inner product space $V$, the inequality set (\r{q9g8hq1}) is valid if and only if $\Phi^* \subset R_F$.
\l{prop:subset2}
\end{thm}

Consider 
\bqy
f_1(x,y,z) & = & x \l{198fnq} \\
f_2(x,y,z) & = & y \\
f_3(x,y,z) & = & xy - z^2. \l{198fnqa}
\eqy
From (\r{198fnq}) to (\r{198fnqa}), we obtain
\bqy
f_1 \, (\< \bfu, \bfu \>, \< \bfv, \bfv \>, \< \bfu, \bfv \>) & = & \< \bfu, \bfu \> \ \ \ge \ \ 0 \l{8v0hwdf} \\
f_2 \, (\< \bfu, \bfu \>, \< \bfv, \bfv \>, \< \bfu, \bfv \>) & = & \< \bfv, \bfv \> \ \ \ge \ \ 0 \l{8v0hwdfb} \\
f_3 \, (\< \bfu, \bfu \>, \< \bfv, \bfv \>, \< \bfu, \bfv \>) & = & \< \bfu, \bfu \> \< \bfv, \bfv \> - \< \bfu, \bfv \>^2 \ \ \ge \ \ 0 \l{8v0hwdfc}
\eqy
respectively, which hold for all $\bfu, \bfv \in V$. 
Note that (\r{8v0hwdf}) and (\r{8v0hwdfb}) are implied by the positive-definiteness
of an inner product space.
From (\r{8v0hwdf}) to (\r{8v0hwdfc}), we see that 
\[
f_i \, (\< \bfu, \bfu \>, \< \bfv, \bfv \>, \< \bfu, \bfv \>) \ge 0
\]
is valid for $i=1,2,3$.
Now define region
\[
\Phi= \left\{ \ (x, y, z) \in \R^3 : x, y \ge 0 \ \mbox{and} \ z^2 \le xy \ \right\}
\]
which in fact is the region $R_F$ with 
$F = \left\{ f_i \ge 0, i = 1, 2, 3 \right\}$. 
Thus by Theorem~\r{prop:subset2}, we have
$\Phi^* \subset \Phi$.

Ultimately, we are interested in obtaining a complete characterization of the achievable region  $\Phi^*$ instead of just an outer bound on it. The question is whether $\Phi$ is indeed equal to $\Phi^*$. If so, we say that the \CS \ is {\em tight}. Otherwise, additional inequalities on the quantities $\< \bfu, \bfu \>$, $\< \bfv, \bfv \>$, and $\< \bfu, \bfv \>$ are needed to completely characterize
$\Phi^*$. 
Such an inequality, if exists, would play the same fundamental role as the \CS.
The next theorem asserts that there exists no constraint on 
$\< \bfu, \bfu \>$, $\< \bfv, \bfv \>$, and $\< \bfu, \bfv \>$ other than 
the \CS \ and positive-definiteness if $\dim(V) \ge 2$, but is not so if $\dim(V) = 0, 1$. We also say that the \CS \ is sharp if and only if $\dim(V) \ge 2$ (by regarding positive-definitness as given).\footnote{
In the literature there have been works on sharpening (or refining)
the \CS, for example \cite{Hovenier93}\cite{Mercer09}. The bounds on $\< \bfu, \bfv \>$ obtained in these works are tighter than the \CS. However, these bounds depend on quantities other than $\< \bfu, \bfu \>$ and $\< \bfv, \bfv \>$, and so they are beyond the scope of the current work. Nevertheless, for 
such a bound, a geometrical framework similar to the one discussed in this section can be introduced, but it would be more complicated because more than three quantities are involved.

Generally speaking, if more information about $\bfu$ and $\bfv$ is available, then tighter bounds on $\< \bfu, \bfv \>$ can be obtained. In particular, if full information about $\bfu$ and $\bfv$ is available, i.e., both $\bfu$ and $\bfv$ are known, then $\< \bfu, \bfv \>$ can be determined exactly.
}

\begin{thm}
For an inner product space $V$, $\Phi^* = \Phi$ if and only if $\dim(V) \ge 2$. For $\dim(V) = 0$, $\Phi^* = \{ 0 \}$, and
for $\dim(V) = 1$, 
\be
\Phi^*= \left\{ \ (x, y, z) \in \R^3 : x, y \ge 0 \ \mbox{and} \ z^2 = xy \ \right\} .
\l{ferwiot}
\ee

\l{thm:main}
\end{thm}

The following proposition is instrumental in proving Theorem~\r{thm:main}.

\begin{prop}
Let $V$ be an inner product space. If $\dim(V) \ge 2$, then for any non-zero vector $\bfu \in V$, there exists a unit vector $\bfw \in V$ such that $\< \bfu, \bfw \> = 0$. 
%If $\dim(V) = 0$ or 1, the same does not hold.
\l{prop:perp}
\end{prop}

\noindent
\proof \ 
Consider an inner product space $V$ with $\dim(V) \ge 2$ and let $\bfu$ be a non-zero vector in $V$. 
Since $\dim(V) \ge 2$, there exists another 
vector $\bft \in V$ such that $\bfu$ and $\bft$ are linearly independent. Let
\[
\bfv = \bft - \frac{\< \bfu, \bft \>}{\< \bfu, \bfu \>} \bfu.
\]
Note that $\< \bfu, \bfu \> > 0$ since $\bfu \ne 0$.
Then
\ba
\< \bfu, \bfv \> 
& = & \biggr\langle \bfu, \bft - \frac{\< \bfu, \bft \>}{\< \bfu, \bfu \>} \, \bfu  \biggr\rangle \\
& = & \< \bfu, \bft \> - \frac{\< \bfu, \bft \> \< \bfu, \bfu \>}{\< \bfu, \bfu \>} \\
& = & 0,
\ea
where the last step is justified because 
$\< \bfu, \bfu \> \ne 0$.
Also, we have
\ba
\< \bfv, \bfv \>
& = & \biggr\langle \bft - \frac{\< \bfu, \bft \>}{\< \bfu, \bfu \>} \bfu , \, \bft - \frac{\< \bfu, \bft \>}{\< \bfu, \bfu \>} \bfu \biggr\rangle \\
& = & \< \bft, \bft \> - 2 \, \frac{\< \bfu, \bft \>^2}{\< \bfu, \bfu \>} + \frac{\< \bfu, \bft \>^2 \< \bfu, \bfu \>}{\< \bfu, \bfu \>^2} \\
& = & \< \bft, \bft \> - \frac{\< \bfu, \bft \>^2}{\< \bfu, \bfu \>} \\
& \ge & 0,
\ea
where the inequality above follows from the \CS \ (\r{Cauchy}) which is tight if and only if 
$\bfu$ and $\bft$ are linearly dependent. Since the latter does not hold by our assumption, we conclude that
$\< \bfv, \bfv \> > 0$, i.e., $\bfv$ is not the zero 
vector. Finally, the proposition is proved by letting
\[
\bfw = \frac{\bfv}{\sqrt{\< \bfv, \bfv \>}} ,
\]
so that 
\[
\< \bfu, \bfw \>
= \frac{ \< \bfu, \bfv \> }{\sqrt{ \< \bfv, \bfv \> }}
= 0
\]
and
\[
\sqrt{\< \bfw, \bfw \>} 
= \sqrt{\frac{ \< \bfv, \bfv \>}{ \< \bfv, \bfv \> }} = 1.
\]
\endpf

%dim(V) \ge 2$. 
%ots, d$ form an orthonormal basis of $V$.
%te 
%
%
%
%
%$ such that $\< \bfu, \bfv \> = 0$, which can be written as
%
%
%
%
%
%
%d b_j \, \bft_j \ \>
%
%, \bft_j \> & = & 0 \nn \\
% & 0 \l{fas8ho} \\
%
%
%use $\bft_i, i = 1, 2, \cdots, d$ form an orthogonal basis of $V$. 
%tion set of (\r{qfupnwe}) (with the $b_k$'s as unknowns) is a $(d-1)$-dimensional subspace of $\mathbb{R}^d$.
% assumption $d \ge 2$ that $\dim(U) = 
%\cdots, b_d$, with $b_k \ne 0$ for at least one $k$,
%
%
%
%
%
%) that $\la \bfv, \bfv \ra = 1$. The proposition is proved.
%\endpf

\bigskip
\noindent
{\em Proof of Theorem~\r{thm:main}.} \
For the case $\dim(V) = 0$, since $V = \{ 0 \}$, we have $\Phi^* = \{ 0 \} \subsetneq \Phi$. 

For the case $\dim(V) = 1$, since for any two vectors $\bfu, \bfv \in V$, one of them is always a scalar multiple of the other, the inequality (\r{Cauchy}) is always satisfied with equality. This implies that 
\be
\Phi^* \subset \left\{ \ (x, y, z) \in \R^3 : x, y \ge 0 \ \mbox{and} \ z^2 = xy \ \right\} ,
\l{qgpvne}
\ee
which is a proper subset of $\Phi$. To prove that 
\[
\left\{ \ (x, y, z) \in \R^3 : x, y \ge 0 \ \mbox{and} \ z^2 = xy \ \right\}
\subset \Phi^* ,
\]
it suffices to show that every $(x,y,z)$ satisfying $x, y \ge 0$ and $z^2 = xy$,
there exist $\bfu, \bfv \in V$ such that 
\be
(x, y, z) = (\< \bfu, \bfu \>, \< \bfv, \bfv \>, \< \bfu, \bfv \> ).
\l{5q9igpn}
\ee

We first consider the case that either $x=0$ or $y=0$. If $x=0$, then $z^2 \le xy$ becomes $z^2 \le 0$, which implies $z=0$. Then (\r{5q9igpn}) is satisfied by letting $\bfu = 0$ and $\bfv \in V$ such that $\< \bfv, \bfv \> = y$. Likewise if $y = 0$.

Now consider the case that $x, y > 0$. Then $z^2 = xy > 0$, which implies
that $z \ne 0$. 
Let $\bfu \in V$ such that 
\be
\< \bfu, \bfu \> = x.
\l{98hq5}
\ee
We need to consider two cases for $z$, namely $z > 0$ and $z < 0$.
First consider the case $ z > 0$. Together with $z^2 = xy$, we have $z = \sqrt{xy}$.
Let $b = \sqrt{y/x}$, so that $b^2 x = y$. Let $\bfv = b \bfu$. Then
\be
\< \bfv, \bfv \> 
= \< b \bfu , b \bfu \> 
= b^2 \< \bfu , \bfu \>
= b^2 x = y,
\l{qasltha}
\ee
and
\be
\< \bfu, \bfv \> 
= \< \bfu, b \bfu \>
= b \< \bfu, \bfu \> 
= bx 
= \sqrt{xy} = z.
\l{uithw}
\ee
Then we see from (\r{98hq5}) to (\r{uithw}) that 
(\r{5q9igpn}) is satisfied. For the case $z < 0$, we have 
$z = -\sqrt{xy}$. 
Then we let $b = - \sqrt{y/x}$
and repeat the above steps to show that (\r{5q9igpn}) is again satisfied.
Therefore, we have proved (\r{qgpvne}) and hence (\r{ferwiot}).

Now consider the case $\dim(V) \ge 2$. It suffices to show that for any $(x,y,z) \in \Phi$, there exist $\bfu, \bfv \in V$ such that (\r{5q9igpn})
is satisfied.
Then $(x,y,z) \in \Phi^*$, showing that $\Phi \subset \Phi^*$.
 
Consider any $(x,y,z) \in \Phi$. 
Let $\bfu \in V$ such that $\< \bfu, \bfu \> = x$. We seek $\bfv \in V$ such that 
\be
\< \bfv, \bfv \> = y
\l{8q9gh}
\ee
and 
\be
\< \bfu, \bfv \> = z.
\l{8q9ghy}
\ee
We first consider the case that either $x=0$ or $y=0$, which can be proved in 
exactly the same way as we have proved the case for $\dim(V) = 1$. 
%If $x=0$, then $z^2 \le xy$ becomes $z^2 \le 0$, which implies $z=0$. Then let $\bfu = 0$ and choose any $\bfv \in V$ such that $\< \bfv, \bfv \> = y$, and we see that (\r{5q9igpn}) is satisfied. Likewise if $y = 0$.
Now consider the case that $x, y > 0$, and choose any $\bfu \in V$ such that $\< \bfu, \bfu \> = x$.
From Proposition~\r{prop:perp}, there exists a unit vector $\bfw \in V$ 
such that $\< \bfu, \bfw \> = 0$.
Let
\[
\bfv = \frac{z}{x} \,
\bfu + \sqrt{ \ y - \frac{z^2}{x}} \, \bfw .
\]
Note that the quantity inside the square root is nonnegative because $z^2 \le xy$.
Since $\< \bfu, \bfw \> = 0$, we have 
\ba
\< \bfv, \bfv \>
= \frac{z^2}{x^2} \< \bfu, \bfu \> + \left( y - \frac{z^2}{x} \right) 
= \left( \frac{z^2}{x^2} \right) x + \left( y - \frac{z^2}{x} \right) 
= \frac{z^2}{x} + \left( y - \frac{z^2}{x} \right)
= y,
\ea
and
\[
\< \bfu, \bfv \> 
= \, \, \biggr\langle \bfu, \frac{z}{x} 
\bfu \biggr\rangle = \frac{z}{x} \< \bfu, \bfu \> = z.
\]
Thus (\r{8q9gh}) and (\r{8q9ghy}) are satisfied.
The theorem is proved.
\endpf

\medskip
\noindent
{\bf Remarks}
\begin{enumerate}
\item 
From the proof of Theorem~\r{thm:main}, we see that when $\dim(V) = 1$,
$\Phi^*$ is exactly equal to the boundary of $\Phi$.
%the ``lower boundary" of $\Phi$ is 
%the set $\{ \, (x,y,z) \in \R^3 : x, y \ge 0, \, z = 0 \, \}$.
%\item 
%When $\dim(V) = 1$, one may write 
%\[
%\Phi^* = \Phi \cap \left\{ \, (x,y,z) \in \R^3 : z^2 \ge xy \, \right\}
%\]
%and identify 
%\[
%\< \bfu, \bfv \>^2 \ge \< \bfu, \bfu \> \< \bfv, \bfv \>
%\]
%as the additional inequality required for characterizing $\Phi^*$, but this interpretation is a little clumsy.
%Instead, it is more direct to specify $\Phi^*$ as in 
%(\r{qgpvne}). 
\item 
From Theorem~\r{thm:main}, we see that $\Phi^*$ for $\dim(V) = 0$ or 
$\dim(V) = 1$ is a subset of $\Phi^*$ for $\dim(V) \ge 2$. Therefore,
there exists no inequality on the quantities 
$\< \bfu, \bfu \>$, $\< \bfv, \bfv \>$, and $\< \bfu, \bfv \>$ that holds for all inner product space $V$ (regardless of the value of $\dim(V)$) which is not implied by the \CS.
\end{enumerate}

To end this section, we argue that the \CS \ can be regarded as sharp even when
$\dim(V) = 0$ or 1 if we take explicit consideration of the dimension of the vector space. Specifically,
\begin{itemize}
\item 
If $\dim(V) = 0$, then for any $\bfu, \bfv \in V$, we have
$\bfu = \bfv = {\bf 0}$, so that $(\< \bfu, \bfu \>, \< \bfv, \bfv \>, \< \bfu, \bfv \> ) = (0,0,0)$. With this additional constraint, we can refined $\Phi$ to
\[
\Phi_0 = \Phi \cap \{ 0 \} = \{ 0 \},
\]
which is equal to $\Phi^*$.
\item
If $\dim(V) = 1$, for any $\bfu, \bfv \in V$, one of them is always a scalar multiple of the other, and the inequality (\r{Cauchy}) is always satisfied with equality. Then by imposing this additional constraint, we can refine $\Phi$ to 
\[
\Phi_1= \Phi \cap \left\{ \ (x, y, z) \in \R^3 : z^2 = xy \ \right\},
\]
which again is equal to $\Phi^*$.
\end{itemize}

\section{Entropy Inequalities}
\l{sec:entropy}
In the last section, we use the Cauchy-Schwarz inequality
to illustrate how a geometrical formulation can potentially lead to interesting results. 
In this section, we apply the same formality to inequalities on the Shannon entropy, which has led to very fruitful and unexpected results in the past over two  decades. This section is a brief exposition of this subject. The reader is referred to  \cite[Chs.~13-15]{Yeung08} and \cite{Yeung15}\cite{Chan11} for more in-depth discussions.\footnote{A general discussion on the Shannon entropy and related inequalities can be found in a blog by Terence Tao \cite{Tao}.}

In this section, all random variables are discrete. The Shannon entropy (or simply entropy when there is no ambiguity) for a random variable $X$ with probability mass function $p(x)$ is defined as 
\[
H(X) = -\sum_{x \in \calS_X} p(x) \log p(x) ,
\]
where $\calS_X$ denotes the support of $X$.
For a pair of jointly distributed random variables $X$ and $Y$ with probability mass function $p(x,y)$, the entropy
is defined as
\[
H(X,Y) = - \sum_{(x,y) \in \calS_{XY}} p(x,y) \log p(x,y),
\]
where $\calS_{XY}$ denotes the support of $p(x,y)$. The entropy for a finite
number of random variables is defined likewise. The entropy for two or more random variables is often called a {\em joint entropy}, although the distinction between entropy and joint entropy is unnecessary.

In information theory (see \cite{Gallager68}\cite{CoverT91}\cite{Yeung08}), entropy is the fundamental measure of information. In addition to entropy, the following quantities are defined:
\ba
\mbox{\em Mutual Information} & & I(X;Y) = H(X) + H(Y) - H(X,Y) \\
\mbox{\em Conditional Entropy} & & H(X|Y) = H(X,Y) - H(Y) \\
\mbox{\em Conditional Mutual Information} & & I(X;Y|Z) = H(X,Z) + H(Y,Z) - H(X,Y,Z) - H(Z).
\ea
These quantities, collectively called {\em Shannon's information measures}, are  used extensively in coding theorems in information theory problems. 

In this section, inequalities on Shannon's information measures are discussed. These inequalities are the main tool for proving converse coding theorems,
which establish that for a particular communication problem, no coding scheme exists if certain conditions are not satisfied. 
In other words, these inequalities establishes the ``impossibilities" in information theory, and they are sometimes referred to as the ``laws of information theory" \cite{Pippenger86}. 

As we see from the above, all Shannon's information measures can be expressed as a linear combinations of entropies. Therefore, inequalities on Shannon's information measures can be written as inequalities on entropies. For this reason, they are referred to as {\em entropy inequalities}.

In this section, we will not focus on the application of entropy inequalities in proving converse coding theorems. Rather, we will focus on these inequalities themselves. Like what we have done in the previous sections, we first introduce a geometrical framework for entropy inequalities.

Let $[n] = \{ 1, \ldots, n \}$, $\sfN = 2^{[n]}$, and $\Nbar = \sfN \bs \{ \emptyset \}$.
Let $\Theta = \{ X_i, i \in [n] \} $ be a collection of $n$ discrete random variables.
Associated with any collection of $n$ random variables are $k \coloneqq 2^n-1$ joint entropies.
For $\alpha \in \sfN$, write 
$X_\alpha = (X_i, i \in \alpha)$, with the convention that $X_\emptyset$ is a constant.  For example, $X_{\{1,2,3\}}$, or simply
$X_{123}$, denotes $(X_1,X_2,X_3)$.
For a collection $\Theta$ of $n$ random variables, define the set function 
$H_\Theta: \sfN \rightarrow \R$ by
\[
H_\Theta(\alpha) = H(X_\alpha), \ \ \ \alpha \in \sfN ,
\]
with $H_\Theta(\emptyset) = 0$ because $X_\emptyset$ is a constant. 
$H_\Theta$ is called the {\em entropy function} of $\Theta$.

Let ${\cal H}_n$ denote $\R^{k}$, the $k$-dimensional Euclidean space,
with the coordinates labeled by $h_\alpha, \alpha \in \Nbar $.
We call ${\cal H}_n$ the {\em entropy space} for $n$ random variables.
As an example, for $n=3$, the coordinates of $\calH_3$ are labelled by 
\[
h_1, h_2, h_3, h_{12}, h_{13}, h_{23}, h_{123},
\]
where $h_{123}$ denotes $h_{\{1,2,3\}}$, etc.
Then for each collection $\Theta$ of $n$ random variables, $H_\Theta$ can be represented by
a column vector $\bfh^\Theta \in {\cal H}_n$, called the {\em entropy vector} of $\Theta$, whose component corresponding to $\alpha$ is equal 
to $H_\Theta(\alpha)$ for all $\alpha \in \Nbar$.
%${\cal H}_n = \mathop{\hbox{\mit I\kern-.2em R}}\nolimits^{2^n-1}$.
On the other hand, a column vector $\bfh \in {\cal H}_n$ is called {\em entropic}\footnote{
Equivalently, $\bfh \in {\cal H}_n$ is entropic if it is achievable by 
some collection of $n$ random variables.
} if
it is equal to the entropy vector $\bfh^\Theta$ of some collection $\Theta$ of $n$ random variables.

\subsection{Unconstrained and Constrained Entropy Inequalities}
\l{sec:EI}
Like what we have done for the Markov inequality and the \CS, we are motivated to define the region
\[
\Gamma_n^* = \{ \, \bfh \in \R^k : \bfh \ \mbox{is entropic} \, \} .
\]
The region $\Gamma_n^*$ is referred to as the
region of entropy vectors.

An entropy inequality $f(\bfh^\Theta) \ge 0$, where $f : \R^k \ra \R$, is {\em valid} if it holds for all collection $\Theta$ of $n$ \rv s. For example, the inequality
\[
H(X_1) + H(X_2) \ge H(X_1, X_2),
\]
or 
\[
I(X_1; X_2) \ge 0 ,
\]
is valid because it holds for any random variables
$X_1$ and $X_2$; this will be further discussed in Section~\r{sec:ST}.
In the sequel, let $n$ be fixed. 
The following proposition is analogous to Proposition~\r{prop:subset}. Its proof is omitted.

\begin{prop}
A set of entropy inequalities  
$
\{ f_i(\bfh^\Theta) \ge 0, 1 \le i \le m \}
$
is valid if and only if
\[
\Gamma_n^* \subset \{ \, \bfh \in \calH_n : f_i(\bfh) \ge 0, 1 \le i \le m \, \}.
\]
\l{prop:subset_E}
\end{prop}

In information theory, we very often deal with entropy inequalities with certain constraints on the joint distribution for the random variables involved. These are called constrained entropy inequalities, and the constraints on the joint distribution can usually be 
expressed as linear constraints on the entropies. In the sequel, we always assume that the constraints on the entropies are of this form. 
The following are some examples:
\begin{enumerate}
\item 
$X_1$ is a function of $X_2$ if and only if 
\[
H(X_1|X_2) = 0.
\]
\item 
$X_1$ and $X_2$ are independent conditioning on $X_3$ if and only if 
\[
I(X_1;X_2|X_3) = 0.
\]
\item 
The Markov chain $X_1 \leftrightarrow X_2 \leftrightarrow X_3 \leftrightarrow X_4$ holds if and only if
\[
\left\{ \
{
\begin{array}{rcl}
I(X_1;X_3|X_2) & = & 0 \\
I(X_1,X_2 ; X_4 |X_3) & = & 0.
\end{array}
}
\right.
\]
\item
Three random variables $X_1$, $X_2$, and $X_3$ are mutually independent if and only if 
\be
H(X_1, X_2, X_3) = H(X_1) + H(X_2) + H(X_3).
\l{aqiuhr14}
\ee
It is not difficult to show that (\r{aqiuhr14}) is equivalent to
\[
\left\{ \
{
	\begin{array}{rcl}
	I(X_1;X_2) & = & 0 \\
	I(X_2;X_3|X_1) & = & 0 \\
	I(X_1;X_3|X_2) & = & 0.
	\end{array}
}
\right.
\]
\end{enumerate}

Suppose there are $q$ constraints on the entropies given by
\[
Q \bfh = 0,
\]
where $Q$ is a $q \times k$ matrix.
%\footnote{
%The equality constraint (\r{Qh=0}) can be generalized to the inequality 
%constraint $Q \bfh \ge 0$. In this case, we do not assume that $Q$ is full row rank.
%All the discussions in Sections~\r{sec2} and \r{sec3} remain valid under this 
%inequality constraint.
%}
Without loss of generality, we can assume that these $q$ constraints are linearly independent,
so that $Q$ is full row rank.
Let
\be
\Phi = \left\{ \, \bfh \in \calH_n : Q \bfh = 0 \, \right\}.
\l{qgepnf}
\ee
In other words, the $q$ constraints confine $\bfh$ to 
a 
linear subspace $\Phi$ in the entropy space.
The following is the constrained version of Proposition~\r{prop:subset_E}.
\begin{prop}
A set of entropy inequalities  
$
\{ f_i(\bfh^\Theta) \ge 0, 1 \le i \le m \}
$
is valid under the constraint $\Phi$ if and only if
\[
( \, \Gamma_n^* \, \cap \, \Phi \, )  \subset \{ \, \bfh \in \calH_n : f_i(\bfh) \ge 0, 1 \le i \le m \, \}.
\]
\l{prop:subset_Ec}
\end{prop}

We will refer to the inequalities in Proposition~\r{prop:subset_E} as (unconstrained) 
entropy inequalities, and the inequalities in Proposition~\r{prop:subset_Ec} as constrained
entropy inequalities. Note that we can let $\Phi = \calH_n$ when there is no constraint on the
entropies.  In this sense, an unconstrained entropy inequality is a special
case of a constrained entropy inequality.

\subsection{Shannon-Type Inequalities} \l{sec:ST}
It is well known in information theory that all Shannon's information measures are nonnegative. This set of inequalities is collectively called the {\em basic inequalities} in information theory. Specifically, these are inequalities of the form
\begin{enumerate}
\item 
$H(X_\alpha) \ge 0$,
\item
$I(X_\alpha; X_\beta) \ge 0$,
\item 
$H(X_\alpha|X_\gamma) \ge 0$,
\item
$I(X_\alpha; X_\beta | X_\gamma) \ge 0$,
\end{enumerate}
where $\alpha$, $\beta$, and $\gamma$ are disjoint subsets of $\sfN$.
On the other hand, the entropy function satisfies the polymatroidal axioms \cite{Fujishige78}: For any $\delta, \sigma \subset \sfN$,
\begin{enumerate}
\item 
$H_\Theta(\emptyset) = 0$;
\item 
$H_\Theta(\delta) \le H_\Theta(\sigma)$ if $\delta \subset \sigma$; 
\item 
$H_\Theta(\delta) + H_\Theta(\sigma) \ge H_\Theta(\delta \cup \sigma) + H_\Theta(\delta \cap \sigma)$.
\end{enumerate}
It can be shown that the basic inequalities and polymatroid axioms are equivalent (see \cite{Yeung08}[Appendix 14.A]).

%It turns out that for a fixed $n$, the set of basic inequalities is equivalent to a smaller subset called the {\em elemental inequalities}. An elemental inequality has one of the following two forms:
%\begin{enumerate}
%\item
%$H(X_i|X_{[n]-\{i\}}), i \in [n]$;
%\item 
%$I(X_i;X_j|X_K)$, where $i \ne j$ and $K \subset [n] - \{i,j\}$.
%\end{enumerate}
%Note that the definition of an elemental inequality depends on the value of $n$. The total number of elemental inequalities is
%\[
%m = n + \left( \begin{array}{c}
%n \\ 2 \end{array} \right) \ 2^{n-2}.
%\]

The basic inequalities, expressed in terms of the entropies, are linear inequalities in $\calH_n$. Denote this set of inequalities by $G \bfh \ge 0$, where $G$ is an $m \times k$ matrix, and define
\[
\Gamma_n = \{ \, \bfh \in \calH_n : G \bfh \ge 0 \, \}.
\]
Since the basic inequalities always hold, we see from Proposition~\r{prop:subset_E} that 
$\Gamma_n^* \subset \Gamma_n$.

{\em Shannon-type inequalities} are  entropy inequalities that are implied by the basic inequalities. Specifically, an entropy inequality $f(\bfh) \ge 0$ is a Shannon-type inequality if and only if 
\[
\Gamma_n \subset \{ \, \bfh \in \calH_n : f(\bfh) \ge 0 \, \}.
\]
More generally, under the linear constraint $\Phi$ 
(cf.\ (\r{qgepnf})), $f(\bfh) \ge 0$ is a Shannon-type inequality if and only if
\[
( \, \Gamma_n \, \cap \, \Phi \, ) \subset \{ \, \bfh \in \calH_n : f(\bfh) \ge 0 \, \}.
\]
Since $\Gamma_n^* \subset \Gamma_n$, it follows that
\[
( \, \Gamma_n^* \, \cap \, \Phi \, ) \subset \{ \, \bfh \in \calH_n : f(\bfh) \ge 0 \, \}, 
\]
which implies that a Shannon-type inequality is valid.

As mentioned earlier in this section, entropy inequalities are the main tool for proving converse coding theorems in information theory. In fact, Shannon-type inequalities had been all the entropy inequalities that were known until the discovery of {\em non-Shannon-type} inequalities in the late 1990s.

Since $\Gamma_n$ is a polyhedral cone, verification of a {\em linear} Shannon-type inequality can be formulated as a linear programming problem \cite{Yeung97}. ITIP \cite{YeungY} was the first software developed for this purpose, which runs on MATLAB.
Subsequently, variants of ITIP with different additional features have been developed. AITIP \cite{AITIP} can produce a human-readable proof and suggest counterexamples when the inequality to be verified is not Shannon-type. PSITIP \cite{PSITIP} can render proofs for converse coding theorems in network information theory \cite{ElGamalK11}. See \cite{itsoc} for a list of related software. Recently, a symbolic approach to the problem that can drastically speed up the computation has been developed \cite{GuoYG23}\cite{GuoYG25}.
For a general discussion on machine-proving of entropy inequalities, we refer the reader to the tutorial paper \cite{YeungLi21}.

%Since the basic inequalities define the region $\Gamma_n$, in general, if $\Phi$ is a collection of functional dependence and conditional independence constraints, then
%$\Gamma_n \cap \Phi$ is a face of $\Gamma_n$.

\subsection{Beyond Shannon-Type Inequalities}
To our knowledge, \cite{Pippenger86} was the first work in the literature that explicitly asked whether there exists any constraint on the entropy function other than the polymatroidal axioms. The same question was raised in \cite{Yeung91b} in a somewhat different form.
With the geometrical formulation for entropy inequalities described at the beginning of this section, it became reasonable to conjecture the existence of constraints on the entropy function beyond Shannon-type inequalities, because it is not readily provable that $\Gamma^*_n = \Gamma_n$.

Before diving deeper into this subject, we first note that
\begin{enumerate}
\item 
$\Gamma_2^* = \Gamma_2$;
\item
$\Gamma_3^* \ne \Gamma_3$ but $\overline{\Gamma_3^*} = \Gamma_3$, where $\overline{\Gamma_3^*}$ denotes the closure of $\Gamma_3^*$;
\end{enumerate}
While it is straightforward to show that $\Gamma_2^* = \Gamma_2$, the problem is already nontrivial for $n=3$. Specifically, along an extreme direction of $\Gamma_3$, only certain discrete points are entropic, making $\Gamma_3^*$ not closed. However, upon taking the closure of $\Gamma_3^*$, we obtain $\Gamma_3$. 

For $n \ge 4$, the set $\Gamma_n^*$ is very complex and characterization of $\Gamma_n^*$ remains an open problem. 
Nevertheless, the following general properties of $\Gamma_n^*$
are known:
\begin{enumerate}
\item
$\overline{\Gamma_n^*}$ is a convex cone;
\item 
${\rm int} \, (\overline{\Gamma_n^*}) \subset \Gamma_n^*$, where
${\rm int} \, \left( \cdot \right)$ denotes the interior of a set \cite{Matus07}.
\end{enumerate}
Here, ${\rm int} \, ( \overline{\Gamma_n^*} ) \subset \Gamma_n^*$ means that the difference between $\Gamma_n^*$ and $\overline{\Gamma_n^*}$ can only be on the boundary. 
As discussed, $\Gamma_3^*$
and $\overline{\Gamma_3^*}$ differ on an extreme direction of $\overline{\Gamma_3^*}$ (= $\Gamma_3$), which is on the boundary of $\overline{\Gamma_3^*}$.

For 4 random variables, the following constrained entropy inequality was proved
\cite{ZhangY97}: If 
\be
I(X_1; X_2) = I(X_1; X_2|X_3) = 0 ,
\l{iubia}
\ee
then
\be
I(X_3; X_4) \le I(X_3; X_4|X_1) + I(X_3; X_4|X_2).
\l{q9pha}
\ee
%Note that (\r{q9pha}) is equivalent to 
%\ba
%& H(X_1, X_2, X_3) + H(X_1, X_4) + H(X_2, X_4) + H(X_3, X_4) - H(X_1, X_2) - H(X_4) & \\
%& - H(X_1, X_3, X_4) - H(X_2, X_3, X_4) \geq 0 , &
%\ea
%which is the {\em Ingleton inequality} for entropy.
This inequality, referred to in the literature as ZY97, cannot be proved by ITIP and hence is a non-Shannon-type inequality.

It was discussed earlier that along an extreme direction of $\Gamma_3$, only certain discrete points are entropic, while the rest are non-entropic.
In the above, the constraints in (\r{iubia}) together with $\Gamma_4$ define a 13-dimensional face\footnote{
For a convex polytope $P$ in $\calH_n$, a face is any set of the form
$F = P \cap \{ \bfh \in \calH_n : \bfb^\top \bfh = c \}$,
where $\bfb^\top \bfh \le c$ for all $\bfh \in P$.
}
of $\Gamma_4$, and ZY97 asserts that a region on this face is not entropic. However, it is still unclear whether $\ol{\Gamma_4^*}$ is equal to $\Gamma_4$. Shortly after, the following unconstrained non-Shannon-type entropy inequality was discovered \cite{ZhangY98}:
For any random variables $X_1$, $X_2$, $X_3$, and $X_4$,
\be
2I(X_3; X_4) \le I(X_1; X_2) + I(X_1; X_3, X_4)
+3I(X_3; X_4|X_1) + I(X_3; X_4|X_2).
\l{ZY98}
\ee
This inequality, referred to in the literature as ZY98 or the {\em Zhang-Yeung inequality}, shows that $\ol{\Gamma_4^*}$ is a proper subset of $\Gamma_4$.
See Fig.~\r{Gamma_4*} for an illustration. 

Subsequently, many non-Shannon-type inequalities for four or more random variables were discovered \cite{MakarychevMRV02}\cite{Zhang03}\cite{Matus07}\cite{XuWS08}\cite{DoughertyFZ11}. 
In particular, the existence of an infinite class of unconstrained non-Shannon-type inequalities for four random variables was proved, implying that $\Gamma_4^*$ (and more generally $\Gamma_n^*$) is not a pyramid \cite{Matus07}. See \cite{Csirmaz14} for a unifying discussion.

The above are the efforts on characterizing $\ol{\Gamma_n^*}$, in particular $\ol{\Gamma_4^*}$, which remains an open problem. Throughout the years, there have have been efforts on characterizing $\Gamma_4^*$, which is even more difficult. Notable works along this line include \cite{Matus06}\cite{ChenY12}\cite{ChenCB21}\cite{LiuC23}\cite{ChenCB25}. The connection with $\Gamma_n^*$ with conditional independence of random variables will be discussed in Section~\r{sec:prob}.

\begin{figure}[t]
\centering
\includegraphics[height=3in]{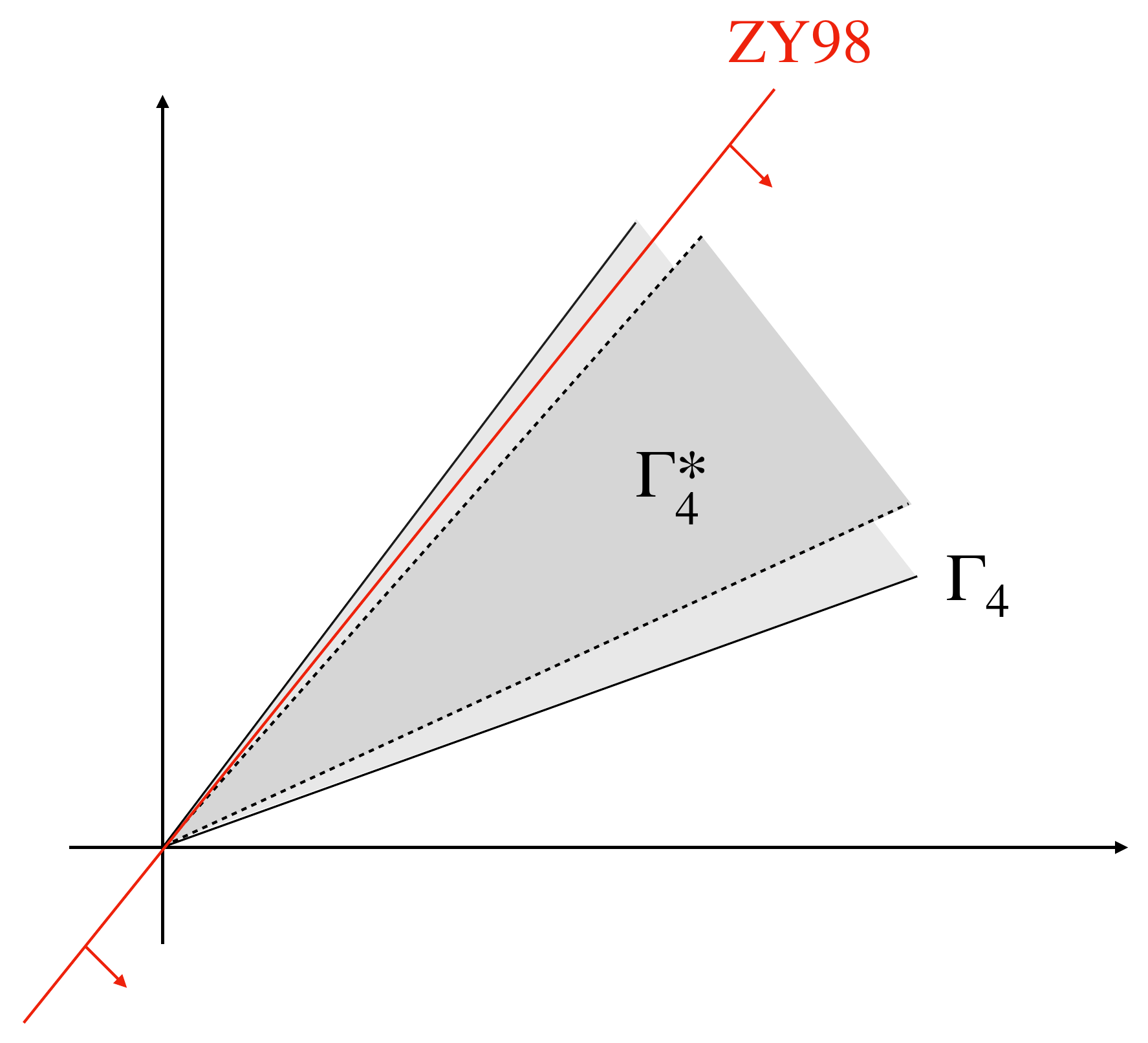}
\caption{An illustration of the non-Shannon-type inequality ZY98.}
\l{Gamma_4*}
\end{figure}

\subsection{Connections with Other Fields}
The study of entropy inequalities, more specifically characterization of the region $\Gamma_n^*$, was shown to be intimately related to a distributed coding problem inspired by satellite communication \cite{YeungZ99b}. This line of research was subsequently
developed into the theory of network coding \cite{AhlswedeCLY00}\cite{YanYZ12}.
In the meantime, intimate relations between this subject and different branches of mathematics and physics were established. In particular, the non-Shannon-type inequalities
for entropy induce corresponding inequalities for finite groups, Kolmogorov complexity, and positive semi-definite matrices. 
In this section, we give a high-level introduction of these developments. For a comprehensive treatment of the topic, we refer the readers to \cite{Yeung15}\cite{Chan11}. 

\subsubsection{Network Coding}
In network communication, to send information from a source node $s$ to a destination node $t$, the predominant existing method is {\em routing}, namely that data packets are routed from node~$s$ to node~$t$
through the intermediate nodes in its original form.
Network coding theory \cite{AhlswedeCLY00} refutes the folklore that routing alone
can achieve the network capacity. Rather, coding at the network nodes, referred to as ``network coding", is in general required. As routing a data packet from an input to an output of a network node can be regarded as applying the identity map to the data packet, routing is a special case of network coding.

\begin{figure}[t]
	\centering
	\includegraphics[height=2.5in]{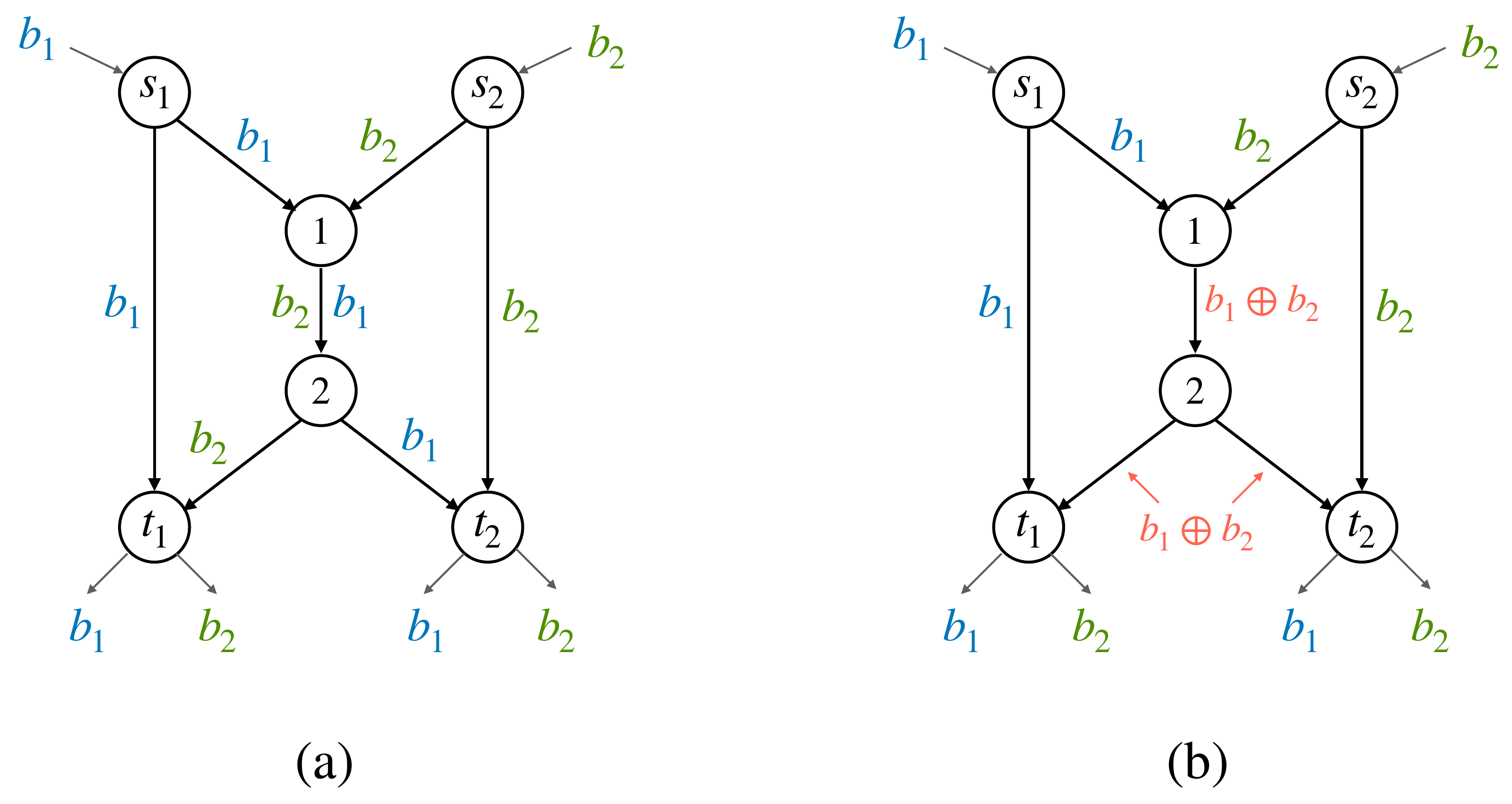}
	\caption{The butterfly network that illustrates the advantage of network coding.}
	\l{butterfly}
\end{figure}

The advantage of network coding can be illustrated by a simple example 
called the {\em butterfly network}, represented by the directed graph in Figure~\r{butterfly}. Here, a directed edge $(i,j)$
represents a communication channel from node~$i$ to node~$j$.
A bit $b_i$ is generated at source node $s_i$, $i = 1, 2$, and the bits $b_1$ and $b_2$ are to be multicast\footnote{In network communication, multicast refers to transmitting a message to a specified subset of destination nodes in the network.} to two destination nodes
$t_1$ and $t_2$. Figure~\r{butterfly}(a) shows a routing solution, in which both $b_1$ and $b_2$ need to be transmitted on channel $(1,2)$. 

If only one bit can be transmitted on each channel, then there exists no routing solution for this multicast problem. However, if an intermediate node can apply computation to the incoming bits instead of just routing them through, then a solution can be obtained as in Figure~\r{butterfly}(b). Here at node~1, the received bits $b_1$ and $b_2$ are combined into a new bit $b_1 \oplus b_2$, where `$\oplus$' denotes
binary addition or exclusive-or (XOR). This operation is referred to as 
{\em network coding}. The bit $b_1 \oplus b_2$ is then transmitted to node~2, where two copies of the bit are sent to the destination nodes $t_1$ and $t_2$, respectively. At node~$t_1$, a copy of the bit $b_1$ is received
directly from node~$s_1$, while the bit $b_2$ can be decoded by adding the two received bits:
\[
b_1 \oplus (b_1 \oplus b_2) = (b_1 \oplus b_1) \oplus b_2 
= 0 \oplus b_2 = b_2 .
\]
Similarly, the bits $b_1$ and $b_2$ can be decoded at node~$t_2$.

Among the vast literature of network coding, \cite{YeungZ99b}, \cite{SongYC06}, and \cite{YanYZ12}
are directly related to the discussions in this section. In a nutshell, these 
works give a complete characterization of the {\em capacity region}\footnote{
The capacity region contains all the achievable information rate tuples of the network coding problem.
} of the general network coding problem 
on an acyclic network in terms of $\Gamma^*$, the region of entropy vectors.
Here, we omit the subscript $n$ in $\Gamma^*$ because the exact number of random variables involved depends on the setup of the specific network coding problem.
Evidently, this characterization is implicit because the complete characterization of $\Gamma^*$ is still open. Nevertheless, an outer (inner) bound on $\Gamma^*$ directly induces an outer (inner) bound on the capacity region of the network coding problem.
For a comprehensive discussion on this topic, we refer the reader to
\cite[Ch.~21]{Yeung08}\cite{YanYZ12}.

\subsubsection{Probability Theory}
\l{sec:prob}
We use $X_\alpha \perp X_\beta \, | \, X_\gamma$ to denote
the conditional independency (CI)
\[
\mbox{$X_\alpha$ and $X_\beta$
	are conditionally independent given $X_\gamma$,}
\]
where $\alpha$, $\beta$, and $\gamma$ are assumed to be disjoint subsets of $[n]$.
%As discussed in Section~\r{sec:EI},
%$X_\alpha \perp X_\beta \, | \, X_\gamma$ is equivalent to
%\be
%I(X_\alpha; X_\beta|X_\gamma) = 0.
%\l{a;b|c}
%\ee
When $\gamma = \emptyset$, $X_\alpha \perp X_\beta \, | \, X_\gamma$
becomes an unconditional independency which we regard as
a special case of a conditional independency.
%When $\alpha = \beta$, (\r{a;b|c}) becomes
%\be
%H(X_\alpha|X_\gamma) = 0,
%\ee
%which we see from Section~\r{sec:EI}  
%that $X_\alpha$ is a function of $X_\gamma$.  For this reason, we also regard
%functional dependency as a special case of conditional independency.

In probability theory, we are often given a set of CI's 
and we need to determine whether another given CI is logically implied.  
We refer to this problem as the {\em implication problem},
which is one of the most basic problems in 
probability theory.
For example, we want to know whether
\[
\left. \begin{array}{l}
X_1 \perp X_3 \, | \, X_2 \\
X_1 \perp X_2
\end{array} \right\}
\Rightarrow X_1 \perp X_3 .
\]
This is not difficult to prove. 
However, the general implication problem is extremely difficult, and it
has been solved only up to four random variables 
\cite{Matus99}.  

We now explain the relation between the implication 
problem and the region $\Gamma_n^*$.
A CI involving random variables 
$X_1, X_2, \cdots, X_n$ has the form 
\be
X_\alpha \perp X_\beta \, | \, X_\gamma,
\l{generic_CI}
\ee
where $\alpha$, $\beta$, and $\gamma$ are disjoint subsets of $[n]$. 
Denote this generic CI by $K$. 
From the discussion in Section~\r{sec:EI}, $K$ is equivalent to 
$I(X_\alpha; X_\beta | X_\gamma) = 0$, i.e., setting the basic inequality
$I(X_\alpha; X_\beta | X_\gamma) \ge 0$ to equality. 
Furthermore, since 
$I(X_\alpha; X_\beta | X_\gamma) = 0$
is equivalent to 
\be
H(X_{\alpha \cup \gamma}) + H(X_{\beta \cup \gamma}) - 
H(X_{\alpha \cup \beta \cup \gamma}) - H(X_\gamma) = 0,
\ee
the CI $K$ corresponds to the following hyperplane in $\calH_n$:
\be
\calE(K) := \left\{ \, {\bf h} \in \calH_n : h_{\alpha \cup \gamma} + h_{\beta \cup \gamma}
- h_{\alpha \cup \beta \cup \gamma} - h_\gamma = 0 \, \right\}.
\ee
Since the region $\Gamma_n$ is defined by the basic inequalities (with $I(X_\alpha; X_\beta | X_\gamma) \ge 0$ being one), $\calE(K) \cap \Gamma_n$ is a face
of $\Gamma_n$.

Let $\Pi = \{ K_l \}$ be a collection of CI's, and we want to 
determine
whether $\Pi$ implies a given CI $K$.  This would be the case if 
and only if the following is true:
\[
\mbox{For all} \ \bfh \in \Gamma_n^*, \ 
\mbox{if} \ \bfh \in \cap_l \, \calE(K_l), \
\mbox{then} \ \bfh \in \calE(K).
\]
Equivalently, 
\[
\Pi \ \mbox{implies} \ K \ \ \ \mbox{if and only if} \ \ \ 
\left[ \cap_l \, \calE(K_l) \right] \cap \Gamma_n^*
\subset \calE(K).
\]
Therefore, the implication problem can be solved if $\Gamma_n^*$ 
can be characterized.  
Since $\Gamma_n^* \subset \Gamma_n$, in the above,
$\left[ \cap_l \, \calE(K_l) \right] \cap \Gamma_n^*$ can be rewritten
as 
\[
\left[ \cap_l \, \calE(K_l) \right] \cap \Gamma_n^* \cap \Gamma_n 
= \cap_l \, \left[ ( \calE(K_l) \cap \Gamma_n ) \cap \Gamma_n^* \right] .
\]
As discussed, $\calE(K_l) \cap \Gamma_n$ is a face of $\Gamma_n$.
In other words, the implication problem can be solved by characterizing $\Gamma_n^*$, in particular characterizing $\Gamma_n^*$ on the faces
of $\Gamma_n$. Therefore, to tackle the implication problem, it is not sufficient just to characterize
$\ol{\Gamma_n^*}$.

Hence, the region $\Gamma_n^*$ is not only of fundamental importance
in information theory, but is also of fundamental importance 
in probability theory. For a more general discussion of this topic, 
we refer the reader to the series of papers \cite{MatusS95}\cite{Matus95}\cite{Matus99}.

\subsubsection{Group Theory}
Let $X_1$ and $X_2$ be any two 
random variables.  Then 
\be
H(X_1) + H(X_2) \ge H(X_1,X_2),
\l{eqn:entropy>}
\ee
which is equivalent to the basic inequality
\be
I(X_1;X_2) \ge 0.
\ee
Let $G$ be any finite group and $G_1$ and $G_2$ be subgroups of 
$G$.
It is well known in group theory\footnote{See for example \cite[Theorem~16.27]{Yeung08} for a proof.} that
\be
|G| \, |G_1 \cap G_2| \geq |G_1| \, |G_2|,
\l{ibnl/bv}
\ee
where $|G|$ denotes the
{\em order} of $G$ and $G_1 \cap G_2$ denotes
the
{\em intersection} of $G_1$ and $G_2$ ($G_1 \cap G_2$ is also
a subgroup of $G$).  
By rearranging the
terms, the above inequality can be written as
\be
\log \frac{|G|}{|G_1|} + \log \frac{|G|}{|G_2|} \ge 
\log \frac{|G|}{|G_1 \cap G_2|}.
\l{eqn:group>}
\ee
By comparing (\r{eqn:entropy>}) and (\r{eqn:group>}), one can easily
identify the one-to-one correspondence between the forms of these two inequalities,
namely that $X_i$ corresponds to $G_i$, $i = 1,2$, and
$(X_1,X_2)$ corresponds to $G_1 \cap G_2$.
While (\r{eqn:entropy>}) is true for any pair of
random variables $X_1$ and $X_2$,
(\r{eqn:group>}) is true for any finite group $G$ and subgroups $G_1$
and $G_2$. As a further example, from the entropy inequality 
\be
H(X_1, X_3) + H(X_2, X_3) \ge H(X_1, X_2, X_3) + H(X_3)
\l{aioqrqr4}
\ee
which is equivalent to $I(X_1; X_2|X_3) \ge 0$, we can obtain the 
group inequality
\be
\log \frac{|G|}{|G_1 \cap G_3|} + \log \frac{|G|}{|G_2 \cap G_3|} 
\ge \log \frac{|G|}{|G_1 \cap G_2 \cap G_3|} + \log \frac{|G|}{|G_3|} 
\l{q9p8ha}
\ee
that holds for all finte group $|G|$ and subgroups $G_1$, $G_2$, and 
$G_3$.

This one-to-one correspondence can be extended to any
random variables $X_1, X_2,$ $\cdots, X_n$ and any finite group $G$ and 
its subgroups $G_1, G_2, \cdots, G_n$. 
For example, consider the non-Shannon-type inequality ZY98
which can be written in terms of joint entropies as follows:
\begin{eqnarray*}
	\left. \begin{array}{r} 
		H(X_1) + H(X_1,X_2) + 2 H(X_3) \\
		+ 2 H(X_4) + 4H(X_1,X_3,X_4) \\
		+ H(X_2,X_3,X_4) 
	\end{array} \right\} \le \left\{ 
	\begin{array}{l}
		3H(X_1,X_3) + 3 H(X_1,X_4) \\
		+ 3H(X_3,X_4) + H(X_2,X_3) \\
		+ H(X_2,X_4)
	\end{array} \right. .
\end{eqnarray*}
%\begin{eqnarray*}
%\lefteqn{H(X_1) + H(X_1,X_2) + 2 H(X_3) + 2 H(X_4) 
%+ \ 4H(X_1,X_3,X_4) + H(X_2,X_3,X_4)} \\
%& \le & 3H(X_1,X_3) + 3 H(X_1,X_4) + 3H(X_3,X_4) 
% + H(X_2,X_3) + H(X_2,X_4).
%\end{eqnarray*}
This entropy inequality, which holds for all random variables $X_1, X_2, X_3$
and $X_4$, corresponds to the group inequality
\begin{eqnarray*}
	\left. \begin{array}{r} 
		|G_1 \cap G_3|^3 \, |G_1 \cap G_4|^3 \\
		\cdot \  |G_3 \cap G_4|^3 \, |G_2 \cap G_3| \\
		\cdot \  |G_2 \cap G_4|
	\end{array} \right\} \le \left\{ 
	\begin{array}{l}
		|G_1| \, |G_1 \cap G_2| \, |G_3|^2 \\
		\cdot \  |G_4|^2 \, |G_1 \cap G_3 \cap G_4|^4 \\
		\cdot \  |G_2 \cap G_3 \cap G_4|
	\end{array} \right. ,
\end{eqnarray*}
%\begin{eqnarray*}
%\lefteqn{|G_1 \cap G_3|^3 \cdot |G_1 \cap G_4|^3 \cdot |G_3 \cap G_4|^3 \cdot |G_2 \cap G_3| \cdot 
%|G_2 \cap G_4|} \\ 
%& \le & |G_1| \cdot |G_1 \cap G_2| \cdot |G_3|^2 \cdot |G_4|^2 \cdot |G_1 \cap G_3 \cap G_4|^4 \cdot 
%|G_2 \cap G_3 \cap G_4| ,
%\end{eqnarray*}
which holds for all finite group $|G|$ and subgroups $G_1$, $G_2$, $G_3$, and 
$G_4$.
We call such an inequality a ``non-Shannon-type" group inequality.
Curiously, there has not been a proof of this inequality based on group theory alone
(without going through the entropy function),
which can shed light on the group-theoretic meaning of this inequality.
Likewise, from any other non-Shannon-type 
entropy inequality, one can obtain the corresponding group inequality.

In the above, we have discussed how to obtain a group inenquality from an entropy inequality. On the other hand,
if a group inequality of the form (\r{eqn:group>}) or (\r{q9p8ha})
holds, then the corresponding entropy inequality of the form (\r{eqn:entropy>}) or
(\r{aioqrqr4}) also holds.

This one-to-one correspondence between entropy inequalities and group inequalities
is intimately related to a combinatorial structure known as the {\em quasi-uniform}
array \cite{Chan01}. This combinatorial structure, inspired by the fundamental notion of {\em strong typicality} in 
information theory, is exhibited by any finite group
and its subgroups. We refer the reader to \cite{ChanY02}\cite[Ch.~16]{Yeung08} for the details.

\subsubsection{Matrix Theory}
\l{matrix}
Let $X$ be a continuous random variable with probability density function (pdf) $f(x)$.
The differential entropy of $X$ is defined as 
\[
h(X) = - \int f(x) \log f(x) dx .
\]
Likewise, the joint differential entropy of a random vector $\bfX$ with joint pdf $f(\bfx)$ is defined as
\be
h(\bfX) = - \int f(\bfx) \log f(\bfx) d \bfx .
\l{hG}
\ee
The integral in the above definitions are assumed to be taken over the support of the underlying
pdf.  

A linear differential entropy inequality
\[
\sum_{\alpha \in \Nbar} c_\alpha h(X_\alpha) \ge 0
\]
is said to be balanced if for all $i \in [n]$, we have $\sum_{\alpha \in \Nbar : i \in \alpha} c_\alpha = 0$.
(The same can be defined for an entropy inequality.) It was proved in \cite{Chan03}  that the above differential entropy inequality 
is valid if and only if it is balanced and its discrete analog is valid.
For example,
\[
h(X|Y) = h(X,Y) - h(Y) \ge 0
\]
is not valid because it is not balanced.  On the other hand,
\[
I(X;Y) = h(X) + h(Y) - h(X,Y) \ge 0
\]
is valid because it is balanced and its discrete analog
\[
H(X) + H(Y) - H(X,Y) \ge 0
\]
is valid.  Thus if $\Gamma_n^*$ can be determined, then in principle all valid differential entropy inequalities can be determined.

Any $n \times n$ symmetric positive semi-definite matrix $K = \left[ \, k_{ij} \, \right]$ defines 
a Gaussian vector $\bfX = [ \, X_1 \ X_2 \ \cdots \ X_n \, ]$ with covariance matrix $K$.
Substituting the corresponding Gaussian distribution into (\r{hG}), we obtain
\[
h({\bf X}) = {1 \over 2} \log \left[ (2 \pi e)^n |K| \right] ,
\]
where $| \cdot |$ denotes the determinant of a matrix.
For $\alpha \in \Nbar$, let $K_\alpha$ be the submatrix of $K$
at the intersection of the rows and the columns 
of $K$ indexed by $\alpha$, whose determinant $|K_\alpha|$ is called a 
{\em principal minor} of $K$.  Note that $K_\alpha$ is the covariance matrix of the subvector 
$\bfX_\alpha = [ \, X_i : i \in \alpha \, ]$.
Since $\bfX_\alpha$ is also Gaussian, it follows that
\be
h({\bf X}_\alpha) = {1 \over 2} \log \left[ (2 \pi e)^{|\alpha|} |K_\alpha| \right] .
\l{hGa}
\ee

Now consider the independence bound for differential entropy,
\[
h(X_1, X_2, \cdots, X_n) \le \sum_i h(X_i) ,
\]
which is tight if and only if $X_i, i \in [n]$ are mutually independent.
Substituting (\r{hGa}) into the above, we have
\[
\half \log [ (2 \pi e)^n |K| ] \le \sum_i \half \log [ (2 \pi e) K_i ] ,
\]
or
\[
{n \over 2} \log (2 \pi e) + \half \log |K| \le {n \over 2} \log (2 \pi e) + \half \log \prod_i K_i .
\]
Note that those terms involving $\half \log (2 \pi e)$ are cancelled out, because the 
independence bound is a valid differential entropy 
inequality and so it is balanced.  After simplification, we obtain
\[
|K| \le \prod_i K_i ,
\]
namely {\em Hadamard's inequality}, which is tight if and only if 
$X_i, i \in [n]$ are mutually independent, or $k_{ij} = 0$ for all $i \ne j$.

This and similar techniques can been applied to obtain various inequalities on the principal minors of symmetric positive semi-definite matrices \cite[Section~16.8]{CoverT91}. These include a generalization of Hardamad's inequality due to Sz\'{a}sz \cite{Mirsky57} and the Minkowski inequality \cite{Minkowski50}.

For every valid differential entropy inequality, a corresponding inequality involving the 
principal minors of a symmetric positive semi-definite matrix can be 
obtained in this fashion.  It turns out that all non-Shannon-type inequalities for discrete random 
variables discovered so far are balanced, and so they are also valid for differential entropy.
For example, from ZY98 we can obtain
\[
|K_1| |K_{12}| |K_3|^2 |K_4|^2 |K_{134}|^4 |K_{234}|
\le
|K_{13}|^3 |K_{14}|^3 |K_{34}|^3 |K_{23}| |K_{24}| ,
\]
which can be called a ``non-Shannon-type" inequality for $4 \times 4$ positive semi-definite matrix $K$.
It was proved in \cite{ChanGY12} that for $3 \times 3$ 
positive semi-definite matrices, all inequalities involving the principal minors 
can be obtained through the Gaussian distribution as explained.  
%In a related work,
%Hassibi and Shadbakht \cite{HassibiS08} studied the properties of normalized Gaussian
%(differential) entropy functions.

\subsubsection{Kolmogorov Complexity}
Kolmogorov complexity, also known as Kolmogorov-Chatin complexity, is a subfield of computer
science.
The Kolmogorov complexity of a sequence $x$, denoted by $K(x)$, is the length of the shortest description
of the string with respective to a {\em universal description language}.   
Without getting into the details, such a universal description language can be based on 
a computer programming language.
Likewise, the Kolmogorov complexity of a pair of sequences $x$ and $y$ is denoted by
$K(x,y)$.  We refer the reader to \cite{LiV08} for a comprehensive treatment of the subject.

Hammer~{\em et~al}.\ \cite{HammerRSV00} established that all linear inequalities 
that are valid for 
Kolmogorov complexity are also valid
for entropy, and vice versa.  For example, the inequality
\[ H(X_1) + H(X_2) \ge H(X_1, X_2) \]
for any $X_1, X_2$
corresponds to the inequality 
\[ K(x_1) + K(x_2) \ge K(x_1,x_2) \]
for any two sequences $x_1$ and $x_2$.
This establishes a one-to-one
correspondence between entropy and Kolmogorov complexity.
Due to this one-to-one correspondence, ``non-Shannon-type" inequalities for Kolmogorov complexity
can be obtained accordingly.

\subsubsection{Quantum Mechanics}
The von Neumann entropy \cite{vN32} is a generalization of the classical entropy (Shannon entropy) to the field of quantum mechanics.\footnote{We refer the reader to \cite{AC2000} for a comprehensive treatment of quantum information theory.} For any quantum state described by a Hermitian positive semi-definite matrix $\rho$, the von Neumann entropy of $\rho$ is defined as
\begin{eqnarray*}
	S(\rho) = - \mathrm{Tr}(\rho \log \rho).
\end{eqnarray*}
Consider distinct quantum systems $A$ and $B$. The joint system is described by a Hermitian positive semi-definite matrix $\rho_{AB}$. The individual systems are described by $\rho_{A}$ and $\rho_{B}$ which are obtained from $\rho_{AB}$ by taking partial trace.
Consider a fixed $\rho_{AB}$.
We simply use $S(A)$ to denote the entropy of System $A$, i.e., $S(\rho_A)$.
In the following, the same convention applies to other joint or individual systems.
It is well known that 
\begin{eqnarray*}
	|S(A) - S({B})| \leq S({AB}) \leq S(A) + S({B}).
\end{eqnarray*}
The second inequality above is called the {\em subadditivity} for the von Neumann entropy.
The first inequality, called the triangular inequality (also known as the Araki-Lieb inequality \cite{AL70}),
is regarded as the quantum analog of the inequality \begin{eqnarray}
H(X) \leq H(X, Y) \label{eq:HX|Y>0}
\end{eqnarray}
for the Shannon entropy.
It is important to note that although the Shannon entropy of a joint system is always not less than the Shannon entropy of an individual system as shown in~(\r{eq:HX|Y>0}), this may not be true in quantum systems. It is possible that $S({AB}) = 0$ but $S({A}) > 0$ and  $S({B}) > 0$, for example, when ${AB}$ is a pure entangled state \cite{AC2000}. From this fact, we can see that the quantum world can be quite different from the classical world.

The {\em strong subadditivity} of the von Neumann entropy \cite{LR73a,LR73b} plays the same role as the basic inequalities for the classical entropy. 
For distinct quantum systems $A$, $B$, and $C$, strong subadditivity can be represented by the following two equivalent forms:
\begin{eqnarray*}
	S({A}) + S({B}) & \leq & S( {AC}) + S( {BC})\\
	S( {ABC}) + S({B}) & \leq & S({AB}) + S( {BC}).
\end{eqnarray*}
These inequalities can be used to show many other interesting inequalities involving conditional entropy and mutual information.
Similar to classical information theory, quantum conditional entropy and quantum mutual information are defined as $S(A|B) = S(A, B) - S(B)$ and $S(A:B) = S(A) + S(B) - S(A, B)$, respectively. For distinct quantum systems $A$, $B$, $C$ and $D$, we have \cite{AC2000}\\
%\begin{enumerate}
i) {\em Conditioning reduces conditional entropy}: 
\begin{eqnarray*}
	S(A|B,C) \leq S(A|B).
\end{eqnarray*}
ii) {\em Discarding quantum systems never increases mutual information}:
\begin{eqnarray*}
	S(A:B) \leq S(A:B,C).
\end{eqnarray*}
iii) {\em Subadditivity of conditional entropy} \cite{Nie98}:
\begin{eqnarray*}
	S(A, B|C, D) & \leq & S(A|C) + S(B|D)\\
	S(A,B|C) & \leq & S(A|C) + S(B|C)\\
	S(A|B,C) & \leq & S(A|B) + S(A|C).
\end{eqnarray*}

Following the discovery of non-Shannon-type inequalities for the classical entropy,
it became natural to ask whether there exist constraints on the von Neumann entropy 
beyond strong subadditivity.
It was proved a few years later that for a three-party system, there exist no such 
constraint \cite{Pippenger03}.  Subsequently, a constrained inequality for the von Neumann entropy for a four-party system which is independent of strong subadditivity was discovered \cite{LW05}, and a family of countably infinitely many constrained inequalities that are independent of each other and strong subadditivity was proved \cite{CLW12}.

%Since $\overline{\Gamma_3^*} = \Gamma_3$, the question is whether $\overline{\Gamma_n^*} = \Gamma_n$ for $n \ge 4$. If so, then Shannon-type inequalities are all entropy inequalities that exist (without imposing conditions on the underlying probability distribution). On the contrary, if 

\section{Concluding Remarks}
\l{sec:conclusion}
In this paper, we have developed a framework for
universally quantified inequalities. With its root in
information theory, this framework provides
a geometrical interpretation that captures the very meaning of such inequalities.
With this formality, we have revisited three celebrated inequalities in mathematics, namely the AM-GM inequality, Markov's inequality, and the \CS,
and clarified the related issues. 
To demonstrate the power of this formality, we have discussed its application 
to the study of entropy inequalities that have yielded very fruitful results 
in a number of subjected related to the Shannon entropy. 
Application of this formality to different branches of mathematics can identify situations in which new fundamental inequalities on quantities of interest may exist, and potentially lead to the discovery of such inequalities.

%proposed a geometrical interpretation for 
%universally quantified inequalities that captures the very meaning of 
%such inequalities. Specifically, a universally quantified inequality is an outer bound on the achievable region.
% The ultimate problem is to completely characterize the achievable region
% We have applied this approach to the AM-GM inequality, the Markov inequality and the Cauchy-Schwarz inequality
% Results not really surprising but help clarify the issues
% This geometrical framework had led to very fruitful results in a number of subjects related to the Shannon entropy
% In particular, non-Shannon-type entropy inequalities were discovered
% Applications to other branches of mathematics may lead to the discovery of new fundamental inequalities on quantities of interest

%\bigskip
%\bigskip
%\noindent
%{\Large \bf Acknowledgment} 
%
%\medskip
%\noindent
%CP Kwong, Chandra, Cheuk Ting, Amin, Satyajit, Qi Chen, Siu Wai, Laigang

\newpage

%\begin{figure}[h]
%\vspace{2in}
%\centering
%\includegraphics[height=3in]{Ex2.pdf}
%\caption{The information diagram for %Example~\r{eg2}.}
%\l{figeg2}
%\end{figure}

\end{document}